\documentclass[preprint,12pt,authoryear,a4paper]{elsarticle}

\usepackage[T1]{fontenc}
\usepackage[utf8]{inputenc}
\usepackage{lmodern}
\usepackage{amsmath, amssymb, amsfonts}
\usepackage{bm}
\usepackage{booktabs}
\usepackage{multirow}
\usepackage{algorithm}
\usepackage{algpseudocode}
\usepackage{xr}
\usepackage[english]{babel}
\usepackage{url}
\usepackage[hidelinks]{hyperref}
\journal{Computational Statistics \& Data Analysis}
\biboptions{round,semicolon}
\makeatletter
\let\ps@pprintTitle\ps@plain
\makeatother
\begin{document}
\begin{frontmatter}

\title{A Generalized Bayesian Approach to Clustering with Dimension Reduction}
% The supplied manuscript had no author details. Fill these before submission.
\author[aff1]{Shen Jia\corref{cor1}\fnref{fn1}}
 \ead{SHJIA@tcd.ie}
 \cortext[cor1]{Corresponding author}
 \fntext[fn1]{These authors contributed equally to this work.}
 \affiliation[aff1]{organization={School of Computer Science and Statistics, Trinity College Dublin},
                   city={Dublin}, country={Ireland}}
\author[aff2]{Weipeng Huang\fnref{fn1}}
\affiliation[aff2]{organization= {School of Computer Science and Software Engineering,
Shenzhen University of Information Technology}, city={Shenzhen}, country={China}}
\author[aff2]{Geng Yang}
\author[aff1]{Eoin Gallagher}
\author[aff1]{Tin Lok James Ng}
%\author{}

\begin{abstract}
    Generalized Bayes product partition models provide a probabilistic counterpart to loss-based clustering. They update a prior over partitions through an exponentiated clustering loss rather than a likelihood, so allocation uncertainty can be quantified without a full generative model for the data. However, their performance depends on the space in which the loss is evaluated. For high-dimensional observations, a squared Euclidean loss on the raw features can be dominated by irrelevant coordinates or fail to reflect cluster structure lying near a nonlinear low-dimensional manifold. We propose an extension in which an autoencoder and the partition are learned together, so that the space in which the clustering loss is evaluated is shaped by the clustering itself. Our generalized posterior combines a reconstruction loss, which keeps the low-dimensional representation faithful to the observed data, with a within-cluster sum of squares in the latent representation, which encourages cluster separation. Posterior computation alternates between Gibbs sampling over partitions and gradient-based updates of the autoencoder network, with the learning rate estimated jointly. We illustrate the approach through simulation studies and applications to four benchmark datasets.
\end{abstract}

% Suggested keywords based on the supplied manuscript; edit as appropriate.
\end{frontmatter}

\newpage
\section{Introduction}
\label{sec:introduction}
\citet{rigon2023generalized} proposed a generalized Bayes framework for probabilistic clustering, based on a class of generalized Bayes product partition models. Following the general principle of \citet{bissiri2016general}, a prior distribution over partitions is updated through an exponentiated negative clustering loss, rather than through a likelihood for the observed data. This construction has several attractive features. First, it does not require a complete probabilistic model for the observations which reduces the risk of misspecification in model based clustering approaches. Second, the clustering discrepancy can be chosen to match the structure of the data and the scientific objective. Furthermore, since the resulting Gibbs or generalized posterior is a probability distribution over partitions, uncertainty in cluster allocation can therefore be summarized by pairwise co-clustering probabilities and partition-level credible sets \citep{wade2018bayesian}.
The performance of the generalized Bayes framework depends crucially on the choice of discrepancy. In the applications considered by \citet{rigon2023generalized}, the loss is evaluated in the space of the observed features. When observations are recorded on hundreds or thousands of features, as for images or sensor signals, this is problematic for several reasons. Irrelevant or redundant coordinates can obscure cluster-relevant structure, and the contrast among pairwise distances deteriorates as dimension grows, making proximity less informative for clustering \citep{aggarwal2001surprising,steinbach2004challenges}. Cluster-relevant variation may also be concentrated near a nonlinear low-dimensional manifold that is poorly represented by distances in the ambient space \citep{tenenbaum2000global}. Simply avoiding a likelihood does not resolve these challenges: a squared Euclidean loss on the raw features is still subject to the same limitations. Designing a more appropriate discrepancy by hand is also difficult when the underlying structure is unknown. One strategy is to reduce dimension first, for example with principal components or a pretrained autoencoder \citep{hinton2006reducing}, and then apply generalized Bayes clustering to the reduced features. Because such a representation is fitted without reference to the clustering criterion, however, it may retain variation that is useful for reconstruction but not for separating clusters.

To address these challenges, we propose \textit{GB Cluster--Joint}, an extension of the generalized Bayes clustering framework of \citet{rigon2023generalized}. The key idea is to evaluate the clustering loss in a learned low-dimensional representation, with the representation and the clustering partition learned jointly. Specifically, an autoencoder maps each observation to a latent vector, while the generalized Bayes update is based on a composite loss with two components. The first is a reconstruction loss in the input space, which encourages the latent representation to retain information about the observed features. The second is a squared Euclidean loss in the latent space, which encourages the representation to capture a geometry that distinguishes between clusters. We develop a computational scheme that couples neural-network training with Gibbs sampling over partitions.

The proposed method is also related to joint deep-clustering methods, which let a clustering objective shape a learned representation. Examples include Deep Embedded Clustering \citep{xie2016dec} and joint nonlinear dimension reduction with $K$-means \citep{yang2017kmeansfriendly}; see \citet{zhou2025survey} for a survey. These methods are formulated as optimization procedures and return a fitted partition or observation-level soft assignments rather than a distribution over partitions. Variational Deep Embedding \citep{jiang2017} does yield probabilistic cluster assignments, but it does so by embedding a Gaussian mixture within a variational autoencoder and hence requires a full generative model for the high-dimensional observations. GB Cluster--Joint instead inherits the loss-based updating of \citet{rigon2023generalized}: it provides a distribution over partitions without specifying a likelihood for the observations.

The remainder of the paper is organized as follows. Section~2 reviews generalized Bayes product partition models. Section~3 introduces GB Cluster--Joint, its computation and the posterior summaries. Sections~4 and~5 report simulation studies and data applications, and Section~6 discusses the main findings and limitations. Throughout, GB Cluster--Raw denotes generalized Bayes clustering applied to the observed features, GB Cluster--TwoStage denotes a reconstruction-only autoencoder followed by generalized Bayes clustering, and GB Cluster--Joint denotes the proposed joint procedure.

\section{Background}
\subsection{Generalized Bayes Product Partition Models}
\label{sec:background_gb_ppm}

Let $\mathbf{X}=(x_1,\dots,x_n)$ denote a collection of observations, where $x_i\in\mathcal{X}\subseteq\mathbb{R}^p$ and $p$ denotes the dimension of the vectorized observations. For example, a $D\times D$ image has dimension $p=D^2$. A clustering of $\mathbf{X}$ is a partition $\mathbf{C}=\{C_1,\dots,C_K\}$ of the index set $\{1,\dots,n\}$ into nonempty, disjoint clusters. It is convenient to encode a partition by an allocation vector $\mathbf{c}=(c_1,\dots,c_n)\in\{1,\dots,K\}^n$, with $C_k(\mathbf{c})=\{i:c_i=k\}$ and $n_k(\mathbf{c})=|C_k(\mathbf{c})|$. Because relabelling the clusters changes $\mathbf{c}$ but not the partition, the inferential target is the partition $\mathbf{C}$ rather than the labels themselves. We write $\mathcal{P}_{n,K}$ for the set of partitions of $\{1,\dots,n\}$ into exactly $K$ nonempty clusters, and $\mathcal{C}_{n,K}=\{\mathbf{c}:\ n_k(c)\ge1,\ k=1,\dots,K\}$ for the corresponding allocation vectors.

Generalized Bayesian inference \citep{bissiri2016general} updates a prior $\pi$ through a loss $\ell$ that measures the incompatibility of a parameter value with the data, rather than through a likelihood. For inverse temperature (learning rate) $\eta>0$ the update is
\begin{equation}\label{eq:gb}
\pi_\eta(\,\cdot\mid \mathbf{X})\propto\pi(\cdot)\exp\{-\eta\,\ell(\,\cdot\,;\mathbf{X})\}.
\end{equation}
When $\ell$ is a negative log-likelihood and $\eta=1$, \eqref{eq:gb} reduces to ordinary Bayesian updating. \citet{rigon2023generalized} applied \eqref{eq:gb} to partitions, using additive clustering losses of the form
\begin{equation}\label{eq:loss}
\ell(\mathbf{c};\mathbf{X})=\sum_{k=1}^K\sum_{i\in C_k(\mathbf{c})}D(x_i;\mathbf{X}_k),\qquad \mathbf{X}_k=\{x_i:i\in C_k\},
\end{equation}
where $D(x_i;\mathbf{X}_k)\ge0$ measures the discrepancy between observation $x_i$ and the cluster to which it is assigned. Combined with a prior $\pi(\mathbf{c})$, the generalized posterior
\begin{equation}\label{eq:gbppm}
\pi_\eta(\mathbf{c}\mid X) \propto \pi(\mathbf{c}) \exp\{-\eta\,\ell(\mathbf{c}; \mathbf{X})\}=\pi(\mathbf{c})\prod_{k=1}^K\prod_{i\in C_k(\mathbf{c})}\exp\{-\eta\,D(x_i;\mathbf{X}_k)\}
\end{equation}
factorizes over clusters.  \citet{rigon2023generalized} fix $K$ and assign a uniform prior on $\mathcal{P}_{n,K}$, so that $\pi(\mathbf{C})=1/S(n,K)$, where $S(n,K)$ is the Stirling number of the second kind. Equivalently, $\pi(\mathbf{c})$ is uniform on $\mathcal{C}_{n,K}$. Since $\mathcal{P}_{n,K}$ is finite, \eqref{eq:gbppm} is a proper distribution whenever the loss is finite.

The special case used in this paper takes $D(x_i;\mathbf{X}_k)=\|x_i-\bar{\mathbf{x}}_k\|_2^2$, where $\bar{\mathbf{x}}_k$ is the mean of the observations in cluster $k$, so that $\ell(c;X)$ is the within-cluster sum of squares. Under the uniform prior, the mode of \eqref{eq:gbppm} is then the $K$-means partition, and the generalized posterior can be viewed as a probabilistic counterpart of $K$-means that also quantifies uncertainty about the allocation.

The learning rate $\eta$ controls how strongly the loss updates the prior. As $\eta\to\infty$, \eqref{eq:gbppm} concentrates on the minimizers of $\ell$; as $\eta \to 0$, it reverts to the prior. The spread of the generalized posterior, and hence the co-clustering probabilities derived from it, therefore depends on $\eta$, whose choice is a recognized challenge in generalized Bayesian inference \citep{bissiri2016general,holmes2017assigning,lyddon2019general,syring2019calibrating}. In Section~\ref{sec:methodology} we treat $\eta$ as unknown and give it a Gamma prior.

\subsection{Autoencoders}\label{sec:background_ae}
An autoencoder \citep{hinton2006reducing} is a neural network trained to reproduce its input
through a low-dimensional representation. It consists of two parametric maps:
an encoder $f_{\boldsymbol{\theta}} : \mathbb{R}^p \to \mathbb{R}^d$, which assigns each observation a latent
representation $ z_i = f_{\boldsymbol{\theta}}( x_i)$, and a decoder
$g_{\boldsymbol{\psi}} : \mathbb{R}^d \to \mathbb{R}^p$, which maps latent vectors back to the input space.
Both maps are typically compositions of affine transformations and elementwise nonlinear activation functions, such as the rectified linear unit.  Fully connected layers are a popular choice for generic vector-valued
inputs, and convolutional layers are commonly used for images. The parameters ${\boldsymbol{\theta}}$ and ${\boldsymbol{\psi}}$
are the weights and biases for the encoder and decoder, respectively. They are typically estimated by minimizing the squared reconstruction error
\begin{equation}\label{eq:recon}
\ell_{\mathrm{recon}}({\boldsymbol{\theta}},{\boldsymbol{\psi}};\mathbf{X})=\sum_{i=1}^n\bigl\|x_i-g_{\boldsymbol{\psi}}(f_{\boldsymbol{\theta}}(x_i))\bigr\|_2^2.
\end{equation}

\section{Methodology}
\label{sec:methodology}

\subsection{Loss and prior specification}

As discussed in Section~\ref{sec:introduction}, evaluating the discrepancy $D(x_i;\mathbf{X}_k)$ of Section~\ref{sec:background_gb_ppm} directly in $\mathbb{R}^p$ can be unreliable when many coordinates are irrelevant or noisy. This motivates defining the clustering discrepancy in a learned lower-dimensional representation. Let $f_{\boldsymbol{\theta}}:\mathbb{R}^p\to\mathbb{R}^d$, $d\ll p$, be the encoder and $g_{\boldsymbol{\psi}}:\mathbb{R}^d\to\mathbb{R}^p$ the decoder of an autoencoder (Section~\ref{sec:background_ae}). Let $z_i=f_{\boldsymbol{\theta}}(x_i)$. The latent centroid of cluster $k$ is
\begin{equation}\label{eq:centroid}
\mu_k(\mathbf{c},{\boldsymbol{\theta}};\mathbf{X})=\frac{1}{n_k(\mathbf{c})}\sum_{i\in C_k(\mathbf{c})} f_{\boldsymbol{\theta}}(x_i),\qquad k=1,\dots,K,
\end{equation}
and, using the squared Euclidean distance, the latent clustering loss is
\begin{equation}\label{eq:loss-cluster}
\ell_\text{cluster}(\mathbf{c},{\boldsymbol{\theta}};\mathbf{X})=\sum_{k=1}^K\sum_{i\in C_k(\mathbf{c})}\bigl\|f_{\boldsymbol{\theta}}(x_i)-\mu_k(\mathbf{c},{\boldsymbol{\theta}};\mathbf{X})\bigr\|_2^2 .
\end{equation}
We add the reconstruction loss
\begin{equation}\label{eq:loss-recon}
\ell_\text{recon}({\boldsymbol{\theta}},{\boldsymbol{\psi}};X)=\sum_{i=1}^n\bigl\|x_i-g_{\boldsymbol{\psi}}\{f_{\boldsymbol{\theta}}(x_i)\}\bigr\|_2^2 ,
\end{equation}
and combine the two as
\begin{equation}\label{eq:composite}
L(\mathbf{c},{\boldsymbol{\theta}},{\boldsymbol{\psi}};\mathbf{X})=\lambda\,\ell_\text{recon}({\boldsymbol{\theta}},{\boldsymbol{\psi}};\mathbf{X})+\ell_\text{cluster}(\mathbf{c},{\boldsymbol{\theta}};\mathbf{X}),
\end{equation}
where $\lambda>0$, calibrated once before posterior computation and then held fixed (as described in Section~\ref{sec:impl}), sets the relative weight of reconstruction. We note that $L$ is a sum of squared residuals over $np$ reconstruction coordinates and $nd$ latent coordinates.

\paragraph{Extended loss function}

The learning rate $\eta$ plays an important role in the generalized posterior. As in \citet{ rigon2023generalized}, we incorporate $\eta$ into the loss function jointly with $(\mathbf{c},{\boldsymbol{\theta}},{\boldsymbol{\psi}})$. Specifically, let $m=n(p+d)/2$ and define the extended composite loss as
\begin{equation}\label{eq:ext-loss}
\ell(c,{\boldsymbol{\theta}},{\boldsymbol{\psi}},\eta;\mathbf{X})=\eta\,L(c,{\boldsymbol{\theta}},{\boldsymbol{\psi}};\mathbf{X})-m \log\eta,\qquad \eta>0 .
\end{equation}
The second term is the loss assigned to $\eta$. It decomposes by component as
\begin{equation}\label{eq:ext-decomp}
\begin{aligned}
\ell(\mathbf{c},{\boldsymbol{\theta}},{\boldsymbol{\psi}},\eta;\mathbf{X})
&=\Bigl\{\eta\lambda\,\ell_\text{recon}({\boldsymbol{\theta}},{\boldsymbol{\psi}};\mathbf{X})-\frac{np}{2}\log\eta\Bigr\}\\
&\quad+\Bigl\{\eta\,\ell_\text{cluster}(\mathbf{c},{\boldsymbol{\theta}};\mathbf{X})-\frac{nd}{2}\log\eta\Bigr\}.
\end{aligned}
\end{equation}
Up to additive constants, the first term is the Gaussian negative log-likelihood of the $np$ reconstruction residuals $x_{ij}-g_{\boldsymbol{\psi}}\{f_{\boldsymbol{\theta}}(x_i)\}_j$ under a common variance $(2\eta\lambda)^{-1}$, and the second is the Gaussian negative log-likelihood of the $nd$ latent residuals $z_{il}-\mu_{c_i,l}$ under a common variance $(2\eta)^{-1}$.

\paragraph{Prior specification and generalized posterior}

For the cluster allocation $\mathbf{c}$ we use the uniform prior $\pi(\mathbf{c})$ on $\mathcal{C}_{n,K}$ of Section~\ref{sec:background_gb_ppm}, which is equivalent to $\pi(\mathbf{C})=1/S(n,K)$ on $\mathcal{P}_{n,K}$. For the autoencoder network parameters we use independent zero-mean Gaussian priors, a standard choice for Bayesian neural networks \citep{neal1996bayesian},
\begin{equation}\label{eq:prior-net}
{\boldsymbol{\theta}}\sim N_{d_{\boldsymbol{\theta}}}(0,\tau_{\boldsymbol{\theta}}^2 I_{d_{\boldsymbol{\theta}}}),\qquad {\boldsymbol{\psi}}\sim N_{d_{\boldsymbol{\psi}}}(0,\tau_{\boldsymbol{\psi}}^2 I_{d_{\boldsymbol{\psi}}}).
\end{equation}
The prior variances shrink weights towards zero, and also identify the scale of the latent representation. Indeed, when the final encoder layer and the first decoder layer are affine, multiplying the encoder's final-layer weights and bias by $s>0$ and the decoder's first-layer weights by $s^{-1}$ leaves $\ell_\text{recon}$ unchanged and multiplies $\ell_\text{cluster}$ by $s^2$. The clustering term therefore rewards shrinking the representation, and the Gaussian prior mitigates it by penalizing large network weights and biases.

Finally, for the learning rate $\eta$ we follow the choice of \cite{rigon2023generalized} and use 
\[\eta\sim\mathrm{Gamma}(a_\eta,b_\eta).\]
Applying the generalized Bayes update of Section~\ref{sec:background_gb_ppm} to the extended loss, with the prior \[ \pi(\mathbf{c},{\boldsymbol{\theta}},{\boldsymbol{\psi}},\eta)=\pi(\mathbf{c})\pi({\boldsymbol{\theta}})\pi({\boldsymbol{\psi}})\pi(\eta),\] gives the generalized posterior
\begin{equation}\label{eq:posterior}
\begin{aligned}
\pi(\mathbf{c},{\boldsymbol{\theta}},{\boldsymbol{\psi}},\eta\mid \mathbf{X}) & \propto \pi(c)\,\pi({\boldsymbol{\theta}})\,\pi({\boldsymbol{\psi}})\,\pi(\eta)\,\eta^m \exp\{-\eta\,L(\mathbf{c},{\boldsymbol{\theta}},{\boldsymbol{\psi}};\mathbf{X})\}.
\end{aligned}
\end{equation}

The full conditional distributions of $\mathbf{c}$ and $\eta$ are available in closed form and underlie the computation in Section~\ref{sec:comp}. Given the other quantities, the allocation has full conditional
\begin{equation}\label{eq:cond-c}
\pi(\mathbf{c}\mid{\boldsymbol{\theta}},{\boldsymbol{\psi}},\eta,X)\propto\pi(\mathbf{c})\exp\{-\eta\,\ell_\text{cluster}(\mathbf{c},{\boldsymbol{\theta}};X)\},\qquad \mathbf{c}\in\mathcal{C}_{n,K},
\end{equation}
because $\ell_\text{recon}$ does not depend on $\mathbf{c}$. The learning rate $\eta$ has full conditional
\begin{equation}\label{eq:cond-eta}
\eta\mid \mathbf{c},{\boldsymbol{\theta}},{\boldsymbol{\psi}},\mathbf{X}\ \sim\ \mathrm{Gamma}\bigl\{a_\eta+ m,  b_\eta+L(\mathbf{c},{\boldsymbol{\theta}},{\boldsymbol{\psi}};\mathbf{X})\bigr\}.
\end{equation}

\subsection{Posterior computation}
\label{sec:comp}
The generalized posterior \eqref{eq:posterior} involves three blocks of unknowns with very different structure, and the computational strategy is designed around those differences. The allocation $\mathbf{c}$ is discrete and combinatorial, with $|\mathcal{C}_{n,K}|$ growing exponentially in $n$, and its components are strongly dependent: whether observation $i$ belongs with $j$ depends on where every other observation is placed. The network parameters $({\boldsymbol{\theta}},{\boldsymbol{\psi}})$ are high-dimensional, and the loss is highly nonconvex in them.  Individual weights lack scientific interpretation; the network matters only through the representation $f_{\boldsymbol{\theta}}$ it induces. The learning rate $\eta$ is a positive scalar.

Exact computation under \eqref{eq:posterior} is intractable. The normalizing constant requires summing over $|\mathcal C_{n,K}|$ allocations and
integrating over the network weights. Sampling-based alternatives for $({\boldsymbol{\theta}},{\boldsymbol{\psi}})$, such as
Hamiltonian Monte Carlo, scale poorly with the number of weights and mix slowly because of the
many symmetries of neural-network parameterizations. We thus use a coordinate-ascent scheme built on a factorized approximation of the generalized posterior in which the network weights are effectively point-estimated, the partition is explored by Gibbs sampling, and $\eta$ is updated in closed form. Specifically, we consider the following approximating family of the generalized posterior:
\[ q(\mathbf{c},{\boldsymbol{\theta}},{\boldsymbol{\psi}},\eta)=q_c(\mathbf{c})\,q_{\boldsymbol{\theta}}({\boldsymbol{\theta}})\,q_{\boldsymbol{\psi}}({\boldsymbol{\psi}})\,q_\eta(\eta)\] with
\begin{equation}\label{eq:families}
\begin{aligned}
q_c &\ \text{unrestricted on } \mathcal{C}_{n,K},\\
q_{\boldsymbol{\theta}}&=N(\hat{{\boldsymbol{\theta}}},\epsilon I),\quad
q_{\boldsymbol{\psi}}=N(\hat{{\boldsymbol{\psi}}},\epsilon I),\\
q_\eta&=\mathrm{Gamma}(a_\eta^q,b_\eta^q),
\end{aligned}
\end{equation}
for a small fixed and arbitrary $\epsilon>0$.  Because $\epsilon$ is fixed, optimizing
over $q_{\boldsymbol{\theta}}$ and $q_{\boldsymbol{\psi}}$ reduces to optimizing their means $(\hat{\boldsymbol{\theta}},\hat{\boldsymbol{\psi}})$. Two considerations motivate this choice. First, the posterior over the weights is not an inferential target. Second, as  $\epsilon\to0$, $q_{\boldsymbol{\theta}}$ and $q_{\boldsymbol{\psi}}$ converge weakly to point masses at $\hat{{\boldsymbol{\theta}}}$ and $\hat{{\boldsymbol{\psi}}}$, which simplifies the update for $(\hat{{\boldsymbol{\theta}}}, \hat{{\boldsymbol{\psi}}})$.

Let $Z$ be the normalizing constant of generalized posterior in \eqref{eq:posterior}. For any $q$ in our approximation family \eqref{eq:families}, Jensen's inequality gives 
\[ \log Z\ge\mathcal{L}(q)\]
with
\begin{equation}\label{eq:elbo}
\mathcal{L}(q)=\mathbb{E}_q\bigl\{\log\pi(\mathbf{c},{\boldsymbol{\theta}},{\boldsymbol{\psi}},\eta)-\log q(\mathbf{c},{\boldsymbol{\theta}},{\boldsymbol{\psi}},\eta)\bigr\}-\mathbb{E}_q\{\ell(c,{\boldsymbol{\theta}},{\boldsymbol{\psi}},\eta;X)\},
\end{equation}
and 
\[\log Z=\mathcal{L}(q)+\mathrm{KL}(q\|\pi).\]
Under the factorization, the expected extended loss is
\begin{equation}\label{eq:exp-loss}
\begin{aligned}
\mathbb{E}_q\{\ell(\mathbf{c},{\boldsymbol{\theta}},{\boldsymbol{\psi}},\eta;X)\}
&=\bar{\eta}\,\mathbb{E}_{q_cq_{\boldsymbol{\theta}} q_{\boldsymbol{\psi}}}\{L(\mathbf{c},{\boldsymbol{\theta}},{\boldsymbol{\psi}};X)\} -m \mathbb{E}_{q_\eta}(\log\eta),\\
\bar{\eta}&=\mathbb{E}_{q_\eta}(\eta)=\frac{a_\eta^q}{b_\eta^q} .
\end{aligned}
\end{equation}

The three factors are coupled  through $\bar\eta$ and the expected loss. The partition
update depends on the network through the current representation, and the network and $\eta$
updates depend on the partition through draws from $q_c$. We maximize \eqref{eq:elbo} one factor at a time while holding the other factors fixed.

\paragraph{Update for $q_c$}
With the other factors fixed, the maximizer over the unrestricted family is
\begin{equation}\label{eq:qc}
q_c^\ast(\mathbf{c})\propto\pi(\mathbf{c})\exp\bigl[-\bar{\eta}\,\mathbb{E}_{q_{\boldsymbol{\theta}} q_{\boldsymbol{\psi}}}\{L(\mathbf{c},{\boldsymbol{\theta}},{\boldsymbol{\psi}};X)\}\bigr]
\approx\pi(\mathbf{c})\exp\{-\bar{\eta}\,\ell_\text{cluster}(\mathbf{c},\hat{{\boldsymbol{\theta}}};\mathbf{X})\},
\end{equation}
using the approximation \[ \mathbb{E}_{q_{\boldsymbol{\theta}} q_{\boldsymbol{\psi}}}\{L\}=L(\mathbf{c},\hat{{\boldsymbol{\theta}}},\hat{{\boldsymbol{\psi}}};\mathbf{X})+O(\epsilon) \approx L(\mathbf{c},\hat{{\boldsymbol{\theta}}},\hat{{\boldsymbol{\psi}}};\mathbf{X})\] 
and dropping $\lambda \ell_\text{recon}$, which does not depend on $\mathbf{c}$. This is the full conditional distribution of $\mathbf{c}$ in \eqref{eq:cond-c} with $({\boldsymbol{\theta}}, {\boldsymbol{\psi}}, \eta)$ taking values $(\hat{{\boldsymbol{\theta}}},\hat{{\boldsymbol{\psi}}},\bar{\eta})$. We sample from $q_c^{*}(\mathbf{c})$ using Gibbs updates.

\paragraph{Update for $(\hat{{\boldsymbol{\theta}}},\hat{{\boldsymbol{\psi}}})$}
With $q_c$ and $q_\eta$ fixed, the terms of \eqref{eq:elbo} depending on $(\hat{{\boldsymbol{\theta}}},\hat{{\boldsymbol{\psi}}})$ are, for small $\epsilon$,
\begin{equation}\label{eq:net-obj}
\mathcal{L}_{{\boldsymbol{\theta}},{\boldsymbol{\psi}}}(\hat{{\boldsymbol{\theta}}},\hat{{\boldsymbol{\psi}}})\approx\log\pi(\hat{{\boldsymbol{\theta}}})+\log\pi(\hat{{\boldsymbol{\psi}}})-\bar{\eta}\,\frac1M\sum_{j=1}^M L\bigl(\mathbf{c}^{(j)},\hat{{\boldsymbol{\theta}}},\hat{{\boldsymbol{\psi}}};\mathbf{X} \bigr),
\end{equation}
where $\mathbf{c}^{(1)},\dots,\mathbf{c}^{(M)}$ are Gibbs draws from $q_c^\ast$. We maximize \eqref{eq:net-obj} by stochastic-gradient ascent.

\paragraph{Update for $q_\eta$}
The terms of \eqref{eq:elbo} involving $\eta$ are $\log\pi(\eta)+ m \log\eta-\eta\,\mathbb{E}_{q_c q_{\boldsymbol{\theta}} q_{\boldsymbol{\psi}}}(L)$, so $q_\eta^\ast$ is Gamma distribution with parameters
\begin{equation}\label{eq:qeta}
\begin{aligned}
a_\eta^q&=a_\eta+m,\\
b_\eta^q&=b_\eta+\mathbb{E}_{q_c q_{\boldsymbol{\theta}} q_{\boldsymbol{\psi}}}(L(\mathbf{c},{\boldsymbol{\theta}},{\boldsymbol{\psi}};\mathbf{X}))\\
&\approx b_\eta+\frac1M\sum_{j=1}^M L\bigl(\mathbf{c}^{(j)},\hat{{\boldsymbol{\theta}}},\hat{{\boldsymbol{\psi}}};\mathbf{X}\bigr),
\end{aligned}
\end{equation}
the conditional \eqref{eq:cond-eta} with $L$ replaced by its expectation.  Since $a_\eta$ and $b_\eta$ are negligible relative to $m$ and $L$, the posterior mean is
$\bar\eta \approx n(p+d)/\{2\,E_q(L)\}$. Since $np \gg nd$, $\bar{\eta}$ is largely determined by how well the autoencoder
reconstructs the data.

The sampling procedure is also related to stochastic-EM-type algorithms \citep{delyon1999convergence}, with a Monte Carlo E-step for the partition via \eqref{eq:qc}, an M-step for the network weights via \eqref{eq:net-obj}, and a closed-form update for $\eta$ via \eqref{eq:qeta}. For posterior inference, we focus on the conditional distribution $\pi(\mathbf{c}\mid\hat{{\boldsymbol{\theta}}},\hat{{\boldsymbol{\psi}}},\bar{\eta},\mathbf{X})$ \eqref{eq:cond-c} evaluated at $( \hat{{\boldsymbol{\theta}}},\hat{{\boldsymbol{\psi}}},\bar{\eta} )$: a plug-in approximation to the generalized posterior over partitions given the fitted representation. Once $(\hat{\boldsymbol{\theta}},\hat{\boldsymbol{\psi}},\bar\eta)$ are fixed, the final Gibbs stage is a standard Markov chain
on the finite space $\mathcal C_{n,K}$ whose stationary distribution is exactly this conditional.
The final draws $\mathbf{c}^{(1)}, \ldots, \mathbf{c}^{(S)}$ do not, however, reflect uncertainty in the fitted representation.

\begin{algorithm}[t]
\caption{Monte Carlo coordinate ascent for GB Cluster--Joint}\label{alg:joint}
\begin{algorithmic}[1]
\Require Data $\mathbf{X}$; number of clusters $K$; priors on $\mathbf{c}$, ${\boldsymbol{\theta}}$, ${\boldsymbol{\psi}}$ and $\eta\sim\mathrm{Gamma}(a_\eta,b_\eta)$; step size of Adam $\alpha$; Gibbs sweeps per iteration $B$; retained draws per iteration $M$; Adam steps per iteration $T$
\State Pretrain $(\hat{{\boldsymbol{\theta}}},\hat{{\boldsymbol{\psi}}})$ on $\ell_\text{recon}$ (warm-up)
\State Initialize $\mathbf{c}$ by $K$-means on $\{f_{\hat{{\boldsymbol{\theta}}}}(x_i)\}$
\State Calibrate $\hat\lambda$ and fix it (Section~\ref{sec:impl})
\State  Set $q_\eta$ by \eqref{eq:qeta} with $M=1$ at the initial $\mathbf{c}$
\For{$t=0,1,2,\dots$ until the stopping rule is met}
  \State $\bar{\eta}\gets a_\eta^q/b_\eta^q$; compute $z_i=f_{\hat{{\boldsymbol{\theta}}}}(x_i)$ and current centroids
  \State \textbf{Sample $\mathbf{c}$:} continuing from the current $\mathbf{c}$, run $B$ Gibbs sweeps using \eqref{eq:qc}; retain $\mathbf{c}^{(1)},\dots,\mathbf{c}^{(M)}$
  \State \textbf{Update $(\hat{{\boldsymbol{\theta}}}, \hat{{\boldsymbol{\psi}}})$:} take $T$ stochastic-gradient ascent steps on \eqref{eq:net-obj} with step size $\alpha$
  \State \textbf{Update $\eta$:} update $(a_\eta^q,b_\eta^q)$ by \eqref{eq:qeta} at the new $(\hat{{\boldsymbol{\theta}}},\hat{{\boldsymbol{\psi}}})$
\EndFor
\State Fix $(\hat{{\boldsymbol{\theta}}},\hat{{\boldsymbol{\psi}}},\bar{\eta})$; run the Gibbs sampler \eqref{eq:qc} for $S$ retained draws
\Ensure $(\hat{{\boldsymbol{\theta}}},\hat{{\boldsymbol{\psi}}})$, $q_\eta$, and partition draws $\mathbf{c}^{(1)},\dots,\mathbf{c}^{(S)}$
\end{algorithmic}
\end{algorithm}

\subsection{Implementation}\label{sec:impl}

\paragraph{Architectures}
For vector-valued inputs, $f_{\boldsymbol{\theta}}$ and $g_{\boldsymbol{\psi}}$ are fully connected networks; for images, convolutional encoders and decoders are used. In all cases the encoder output layer and decoder input layer are affine, and the decoder output layer is linear for standardized continuous inputs. 

\paragraph{Gradient-based calibration of $\lambda$}
After the reconstruction warm-up, $\lambda$ is chosen so that the weighted reconstruction and clustering terms contribute gradients of comparable magnitude to the encoder. At iteration $t$ of a calibration window $\mathcal{T}_{\mathrm{cal}}$, let $G_{\mathrm{rec}}^{(t)}=\|\nabla_{\hat{{\boldsymbol{\theta}}}}\ell_\text{recon}^{(t)}\|_2$ and $G_{\mathrm{cluster}}^{(t)}=\|\nabla_{\hat{{\boldsymbol{\theta}}}}\ell_\text{cluster}^{(t)}\|_2$. For a target ratio $r_0$ and a small $\varepsilon_g>0$,
\begin{equation}\label{eq:lambda}
\hat\lambda=\frac{r_0}{\operatorname{median}_{t\in\mathcal{T}_{\mathrm{cal}}}\bigl\{G_{\mathrm{rec}}^{(t)}/(G_{\mathrm{cluster}}^{(t)}+\varepsilon_g)\bigr\}} .
\end{equation}
We use $r_0=1$ in our experiments and hold $\hat\lambda$ fixed thereafter. We conduct sensitivity analysis with respect to the choice of $r_0\in\{0.5,1,2\}$.

\paragraph{Hyperparameters}
The choice of $a_\eta, b_\eta$ has negligible influence and thus, in all experiments we set $a_\eta=b_\eta=1$.
The encoder and decoder parameters were assigned independent
zero-centred Gaussian priors with
\(\tau_{\boldsymbol{\theta}}^2=\tau_{\boldsymbol{\psi}}^2=10^4\), corresponding to prior
precisions of \(10^{-4}\). At each training iteration, we ran $B=10$ Gibbs sweeps and
retained one partition after each of the final $M=5$ sweeps
to approximate the expected loss.
The final sampling stage continues from the last partition of the joint stage and retains
$S=500$ draws without further burn-in. 
\\\\
Code implementing the proposed method is available at
\url{https://github.com/sj-631/GB-Custer-Joint}.

\subsection{Posterior summaries}\label{sec:summ}

Let $C^{(s)}=C(\mathbf{c}^{(s)})$, $s=1,\dots,S$, be the partitions from the final sampling stage of Algorithm~\ref{alg:joint}. 
We compute a point estimate with SALSO \citep{dahl2022search} under the variation-of-information loss \citep{meila2007comparing}, minimizing the Monte Carlo expected loss $\hat R_{\mathrm{VI},S}(\tilde{C})=S^{-1}\sum_s\mathrm{VI}(\tilde{C},C^{(s)})$ by randomized greedy search. The resulting $\hat{\mathcal{C}}_{\mathrm{SALSO}}$ is used to compute the adjusted Rand index in Sections~4 and~5. As in \citet{rigon2023generalized}, cluster allocation uncertainty is summarized by $\hat P_{ij}=S^{-1}\sum_s\mathbb{I}\{c_i^{(s)}=c_j^{(s)}\}$. We note that all summaries are invariant to relabelling.

\subsection{Selecting the number of clusters}\label{sec:choose-K}
We choose $K$ by fitting GB Cluster--Joint separately at each of several candidate values.
For a given $K$, let $\hat{\boldsymbol{\theta}}$ be the fitted encoder parameters,
$\hat{\mathbf{C}}_{\mathrm{SALSO}}$ the SALSO point estimate of Section~\ref{sec:summ},
$z_i=f_{\hat{\boldsymbol{\theta}}}(x_i)$ and $\bar z=n^{-1}\sum_{i=1}^n z_i$; all of these
depend on $K$, which we suppress. The normalized clustering loss is
\begin{equation}\label{eq:QK}
Q_K=\frac{\ell_\text{cluster}(\hat{\mathbf{C}}_{\mathrm{SALSO}},\hat{\boldsymbol{\theta}};\mathbf{X})}
{\sum_{i=1}^n\|z_i-\bar z\|_2^2},
\end{equation}
where $\ell_\text{cluster}$ in depends on the allocation only
through the partition, we evaluate it directly at $\hat{\mathbf{C}}_{\mathrm{SALSO}}$. The denominator is the total latent sum of squares. Provided
this quantity is positive, $Q_K\in[0,1]$, with smaller values
indicating that a larger proportion of the latent variation
lies between clusters. As in \citet{rigon2023generalized}, we use the elbow
heuristic: the choice of $K$ is guided by the point beyond which further increases in $K$
reduce $Q_K$ only slightly.

\section{Simulation Studies}

We conduct simulation studies to assess whether jointly learning the representation and the partition improves recovery of the generating partition compared with clustering the observed features directly or clustering a representation learned solely through reconstruction. We further investigate the conditions under which the benefits of joint learning are greatest, focusing on cluster separation, ambient dimension, and the number of clusters.  Finally, we evaluate the ability of the elbow criterion to recover the true number of clusters. We compare five methods: $K$-means and GB Cluster--Raw on the observed features; AE + $K$-means and GB Cluster--TwoStage on the embedding of an autoencoder trained on reconstruction alone; and GB Cluster--Joint. The two-stage methods and GB Cluster--Joint use the same architecture, latent dimension and pretraining, so that the comparison of GB Cluster--TwoStage with GB Cluster--Joint isolates the effect of joint training, and the comparison of GB Cluster--Raw with GB Cluster--TwoStage isolates the effect of the representation. Clustering performance is evaluated by comparing the estimated
partition with the true generating labels using the adjusted
Rand index (ARI) \citep{hubert1985comparing}.
For the generalized Bayes methods, we use the SALSO
representative partition (Section~\ref{sec:summ});
for \(K\)-means and AE + \(K\)-means, we use their fitted partitions.
We repeat each configuration five times and report the mean and standard deviation.

\subsection{Designs}\label{sec:sim-design}

\paragraph{Simulation I}
For this first simulation study, we consider the setting where clustering signal is contained in a subset of observed coordinates. We adapt the design from \citet{raymaekers2022regularized} which places the cluster signal directly in 50 observed coordinates and surrounds it with a growing number of pure-noise coordinates. We fix $n=800$ and draw labels $Y_i\sim\mathrm{Categorical}(1/K,\dots,1/K)$. For $k=1,\dots,K$, let $\alpha_k=\pi/4+2\pi(k-1)/K$ and
\begin{equation}\label{eq:sim1-centres}
\nu_k=\sqrt2\,\gamma\begin{pmatrix}\cos(\alpha_k)\,\mathbf 1_{25}\\ \sin(\alpha_k)\,\mathbf 1_{25}\end{pmatrix}\in\mathbb{R}^{50}.
\end{equation}
Given $Y_i=k$, the 50 informative coordinates of $x_i$ are distributed as $N_{50}(\nu_k,I_{50})$, and for $p>50$, they are augmented with $p-50$ independent $N(0,1)$ nuisance coordinates. The distance between adjacent centres is
\begin{equation}\label{eq:sim1-sep}
\delta_{K,\gamma}=10\sqrt2\,\gamma\,\sin(\pi/K),
\end{equation}
so separation increases linearly in $\gamma$ and decreases with $K$. We use all combinations of $K\in\{4,6,8\}$, $p\in\{50,200,500,1000\}$ and $\gamma\in\{0.3,0.4,0.5,0.6\}$, giving a total of 48 configurations.

\paragraph{Simulation II}
For the second simulation study, we adopt a design which places the cluster structure in a
low-dimensional space and observes it only through a nonlinear map with fixed network
parameters. A total of $n=1500$ latent points
$u_i=(u_{i1},\ldots,u_{id_0})^\top\in\mathbb{R}^{d_0}$ are drawn (here $d_0$
denotes the dimension of the data-generating latent space), with component labels
generated independently according to
$Y_i\overset{\text{iid}}{\sim}\mathrm{Categorical}(1/K,\ldots,1/K)$. The first two coordinates
satisfy
\begin{equation}\label{eq:sim2-latent}
\begin{aligned}
(u_{i1},u_{i2})^\top\mid Y_i=k&\sim N_2\bigl(\nu_k,\,0.8^2I_2\bigr),\\
\nu_k&=2.2\bigl(\cos\{2\pi(k-1)/K\},\,\sin\{2\pi(k-1)/K\}\bigr)^\top.
\end{aligned}
\end{equation}
For $d_0>2$, the remaining coordinates are sampled independently from $N(0,0.8^2)$.  The latent points are mapped to $p$ dimensions by the
nonlinear map
\begin{equation}\label{eq:sim2-map}
a_i=\mathrm{ReLU}(W_1u_i+b_1),\qquad x_i=W_2\tilde a_i,
\end{equation}
where the tilde denotes coordinate-wise standardization of the hidden activations $a_i$ over
the sample, $W_1\in\mathbb{R}^{10\times d_0}$, $b_1\in\mathbb{R}^{10}$ and
$W_2\in\mathbb{R}^{p\times 10}$, with entries drawn once and held fixed:
$(W_1)_{jr}\sim N(0,1/2)$, $(b_1)_j\sim\mathrm{Uniform}(-\pi,\pi)$,
$(W_2)_{hj}\sim N(0,1/10)$, where the second argument of $N(\cdot,\cdot)$ denotes the variance.
We consider the reference setting of $K=3$, $d_0=2$ and $p=500$. We then vary one factor at a
time from this reference: the number of clusters $K\in\{3,5,7\}$, where larger $K$ places
adjacent centres closer together; the generating dimension $d_0\in\{2,4,8\}$, where the extra
$d_0-2$ coordinates are label-independent Gaussian noise and the first two latent coordinates
and the corresponding columns of $W_1$ are as in the reference; the ambient dimension
$p\in\{200,500,1000\}$; and the number of affine layers of the map, from two to four. Each
additional hidden layer has width 10, with $N(0,1/10)$ weights, $\mathrm{Uniform}(-\pi,\pi)$
biases and a ReLU activation; the first hidden layer and $W_2$ are shared across depths, and
only the final hidden layer is standardized.

\subsection{Results}\label{sec:sim-results}

\subsubsection{Simulation I}\label{sec:sim1-res}

Results from Simulation I for the case of $K=4$ are shown in Table~\ref{tab:sim-one-ari-n800}, and in Supplementary Tables~S2 and~S3 for $K=6$ and $K=8$, respectively. GB Cluster--Joint had the highest or joint-highest mean ARI in 45 of the 48 configurations. It exceeded GB Cluster--Raw in all 48 configurations and matched or exceeded GB Cluster--TwoStage in 47 configurations.

A clear pattern from Table~\ref{tab:sim-one-ari-n800} is that performance gain from joint training grows with the number of nuisance coordinates. For $K=4$, the average advantage of GB Cluster--Joint over GB Cluster--TwoStage rises from 0.007 at $p=50$, where every coordinate is informative, to 0.081 at $p=1000$, where 95\% are noise. The largest single gains occur at $p=1000$ with weak or moderate separation, for example 0.213 versus 0.128 at $\gamma=0.3$ and 0.682 versus 0.588 at $\gamma=0.4$. We also see that a reconstruction-only representation does not reliably improve performance. For $K=4$, neither GB Cluster--TwoStage nor AE + $K$-means consistently improved on its observed-space counterpart; at $p=1000$, GB Cluster--TwoStage was below GB Cluster--Raw for every $\gamma$. A plausible explanation is that an autoencoder trained to reconstruct all coordinates devotes much of its capacity to the noise block. Performance of all methods deteriorated when weak separation, many nuisance coordinates and many clusters occurred together, and joint training reduced, but did not remove, this deterioration.

\begin{table}[htbp]
    \centering
    \caption{Mean adjusted Rand index (ARI), with sample standard deviations
in parentheses (5 runs with different random seeds) for the
reported $(p,\gamma)$ configurations in Simulation~I with \(K=4\) and
\(n=800\).}
    \label{tab:sim-one-ari-n800}
    \resizebox{\linewidth}{!}{%
    \begin{tabular}{lrrrrr}
        \toprule
        Method & $p$ & $\gamma=0.3$ & $\gamma=0.4$
        & $\gamma=0.5$ & $\gamma=0.6$ \\
        \midrule
        GB Cluster--Raw & 50   & 0.645 (0.031) & 0.884 (0.018) & 0.968 (0.010) & 0.997 (0.001)\\
        GB Cluster--Raw & 200  & 0.558 (0.029) & 0.785 (0.033) & 0.912 (0.017) & 0.993 (0.005) \\
        GB Cluster--Raw & 500  & 0.389 (0.050)& 0.713 (0.033) & 0.924 (0.013)& 0.979 (0.006)\\
        GB Cluster--Raw & 1000 & 0.157 (0.035)& 0.593 (0.026)& 0.856 (0.017) & 0.940 (0.020) \\
        \addlinespace
  K-Means & 50   & \textbf{0.671 (0.027)}  & 0.864 (0.012) & 0.968 (0.010) & 0.995 (0.004) \\
K-Means & 200  & 0.559 (0.039) & 0.820 (0.035) & 0.953 (0.016) & 0.985 (0.005) \\
K-Means & 500  & 0.339 (0.063) & 0.667 (0.088) & 0.887 (0.028) & 0.978 (0.002) \\
K-Means & 1000 & 0.121 (0.028) & 0.573 (0.047) & 0.791 (0.051) & 0.899 (0.052) \\
\addlinespace
        GB Cluster--TwoStage & 50   & 0.642 (0.017)  & 0.893 (0.040)  & 0.970 (0.006)   & \textbf{0.999 (0.001)}   \\
        GB Cluster--TwoStage & 200   & 0.552 (0.020)  & 0.784 (0.022)   & 0.947 (0.012)   & \textbf{1.000 (0.000)}   \\
        GB Cluster--TwoStage & 500  & 0.374 (0.021)  & 0.718 (0.020)   & 0.944 (0.011)  &   0.979 (0.009) \\
        GB Cluster--TwoStage & 1000 & 0.128 (0.049)   &  0.588 (0.029)  & 0.839 (0.057)  &  0.920 (0.009) \\
        \addlinespace
        AE+K-Means & 50   & 0.644 (0.024) & 0.859 (0.041) & 0.964 (0.006) & 0.993 (0.003) \\
        AE+K-Means & 200  & 0.534 (0.031) & 0.801 (0.043) & 0.952 (0.012) & 0.988 (0.004) \\
        AE+K-Means & 500  & 0.352 (0.021) & 0.670 (0.048) & 0.898 (0.025) & 0.963 (0.020) \\
        AE+K-Means & 1000 & 0.116 (0.037) & 0.559 (0.026) & 0.831 (0.056) & 0.911 (0.010) \\
        \addlinespace
        \textit{GB Cluster--Joint} & 50   & 0.648 (0.022) & \textbf{0.900 (0.041)} & \textbf{0.983 (0.011)} & \textbf{0.999 (0.001)} \\
        \textit{GB Cluster--Joint} & 200  & \textbf{0.591 (0.039)} & \textbf{0.849 (0.018)} & \textbf{0.962 (0.014)} & \textbf{1.000 (0.000)} \\
        \textit{GB Cluster--Joint} & 500  & \textbf{0.428 (0.043)} & \textbf{0.777 (0.017)} & \textbf{0.953 (0.012)} & \textbf{0.991 (0.007)} \\
        \textit{GB Cluster--Joint} & 1000 & \textbf{0.213 (0.027)} & \textbf{0.682 (0.038)} & \textbf{0.920 (0.024)} & \textbf{0.983 (0.007)}\\
        \addlinespace
        \bottomrule
    \end{tabular}
    }
    \par\smallskip
\begin{minipage}{\linewidth}
\raggedright
\footnotesize
\textit{Note: All numerical results are mean adjusted Rand indices, with standard
deviations reported in parentheses. Results are summarized over five fitting
seeds. The best result for each setting is shown in bold.}
\end{minipage}
\end{table}

\subsubsection{Simulation II}\label{sec:sim2-res}

For the second simulation study, GB Cluster--Joint had the highest mean ARI in every configuration (Table~\ref{tab:simulation2_results}). In contrast to Simulation~I, representation learning was essential here: in the reference setting, mean ARI was 0.495 for GB Cluster--Raw, 0.853 for GB Cluster--TwoStage and 0.912 for GB Cluster--Joint.  Joint training added 0.036 to 0.079 over GB Cluster--TwoStage and 0.067 to 0.119 over AE + $K$-means across configurations.

Among the factors varied, the number of clusters had by far the largest effect: moving from $K=3$ to $K=7$ lowered the mean ARI of GB Cluster--Joint from 0.912 to 0.375. Adding label-independent generating coordinates ($d_0=2\to8$: $0.912\to0.787$) and deepening the map ($2\to4$ affine layers: $0.912\to0.837$) had moderate effects, while increasing the ambient dimension had the smallest effects ($p=200\to1000$: $0.936\to0.885$). 

\begin{figure}[htbp]
    \centering
    \includegraphics[
        width=\textwidth
    ]{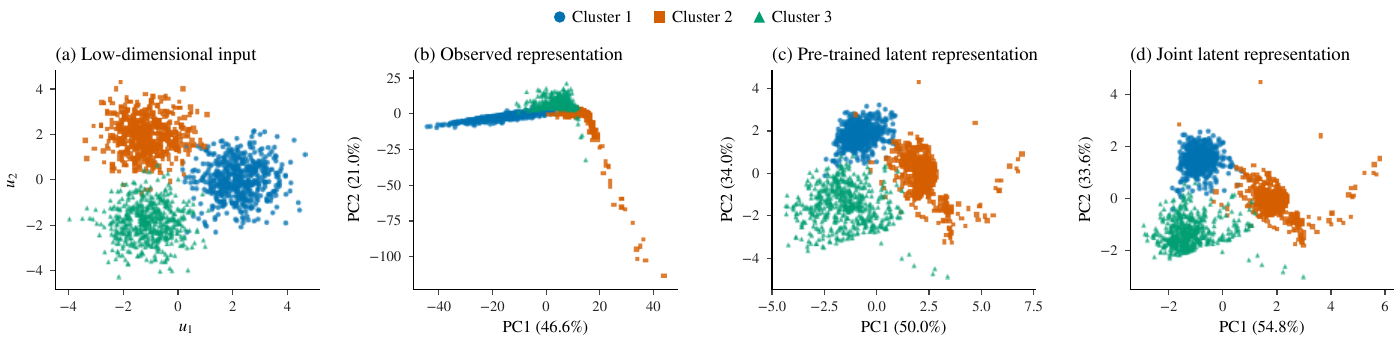}
    \caption{
        Visualization of the reference setting in Simulation II.
        Panel (a) shows the original two-dimensional Gaussian-mixture
        observations.
        Panel (b) shows the first two principal components of the
        500-dimensional observations generated by the fixed nonlinear
        neural-network mapping.
        Panel (c) shows the first two principal components of the
        latent dimension representations learned by the separately
        trained autoencoder. Panel (d) shows the corresponding
        principal-component projection of the representations
        learned by GB Cluster--Joint.
    }
    \label{fig:simulation2}
\end{figure}

Figure~\ref{fig:simulation2} illustrates the reference setting.
The three generating clusters are well separated in the latent
space (panel~a), whereas the PCA projection of the observed data
shows an elongated pattern with overlapping clusters (panel~b).
In the PCA projections of the learned representations,
reconstruction-only training yields partial separation, with
two clusters remaining adjacent (panel~c), whereas joint training
yields more compact and better-separated clusters (panel~d).

\begin{table}[htbp]
\centering
\caption{Summary of the Simulation II settings and clustering performance.}
\label{tab:simulation2_results}
\small

\begingroup
\setlength{\tabcolsep}{3.5pt}
\renewcommand{\arraystretch}{1.08}

\resizebox{\textwidth}{!}{
\begin{tabular}{ccccc rrrrr}
\toprule
$K$
& $d_0$
& $p$
& $Layers$
& \shortstack{Mapping\\architecture}
& \shortstack{$K$-means}
& \shortstack{AE +\\$K$-means}
& \shortstack{GB Cluster\\--Raw}
& \shortstack{GB Cluster\\--TwoStage}
& \shortstack{\textit{GB Cluster}\\\textit{--Joint}} \\
\midrule

\multicolumn{10}{l}{\textit{Varying the number of clusters}} \\
3 & 2 & 500 & 2 & $2\rightarrow10\rightarrow500$
& 0.466 (0.000) & 0.817 (0.040)
& 0.495 (0.001) & 0.853 (0.030) & \textbf{0.912 (0.012)} \\
5 & 2 & 500 & 2 & $2\rightarrow10\rightarrow500$
& 0.245 (0.000) & 0.522 (0.021)
& 0.287 (0.001) & 0.579 (0.037) & \textbf{0.638 (0.020)} \\
7 & 2 & 500 & 2 & $2\rightarrow10\rightarrow500$
& 0.209 (0.000) & 0.308 (0.018)
& 0.208 (0.002) & 0.299 (0.018) & \textbf{0.375 (0.015)} \\

\addlinespace
\multicolumn{10}{l}{\textit{Varying the generating latent dimension}} \\
3 & 2 & 500 & 2 & $2\rightarrow10\rightarrow500$
& 0.466 (0.000) & 0.817 (0.040)
& 0.495 (0.001) & 0.853 (0.030) & \textbf{0.912 (0.012)} \\
3 & 4 & 500 & 2 & $4\rightarrow10\rightarrow500$
& 0.419 (0.000) & 0.796 (0.022)
& 0.493 (0.003) & 0.802 (0.010) & \textbf{0.881 (0.029)} \\
3 & 8 & 500 & 2 & $8\rightarrow10\rightarrow500$
& 0.547 (0.000) & 0.689 (0.031)
& 0.546 (0.001) & 0.711 (0.009) & \textbf{0.787 (0.008)} \\

\addlinespace
\multicolumn{10}{l}{\textit{Varying the observed dimension}} \\
3 & 2 & 200 & 2 & $2\rightarrow10\rightarrow200$
& 0.467 (0.000) & 0.853 (0.023)
& 0.560 (0.005) & 0.890 (0.017) & \textbf{0.936 (0.009)} \\
3 & 2 & 500 & 2 & $2\rightarrow10\rightarrow500$
& 0.466 (0.000) & 0.817 (0.040)
& 0.495 (0.001) & 0.853 (0.030) & \textbf{0.912 (0.012)} \\
3 & 2 & 1000 & 2 & $2\rightarrow10\rightarrow1000$
& 0.462 (0.000) & 0.805 (0.059)
& 0.505 (0.003) & 0.820 (0.022) & \textbf{0.885 (0.013)} \\

\addlinespace
\multicolumn{10}{l}{\textit{Varying the network depth}} \\
3 & 2 & 500 & 2 & $2\rightarrow10\rightarrow500$
& 0.466 (0.000) & 0.817 (0.040)
& 0.495 (0.001) & 0.853 (0.030) & \textbf{0.912 (0.012)} \\
3 & 2 & 500 & 3 & $2\rightarrow10\rightarrow10\rightarrow500$
& 0.431 (0.000) & 0.777 (0.038)
& 0.446 (0.003) & 0.834 (0.025) & \textbf{0.896 (0.019)} \\
3 & 2 & 500 & 4
& $2\rightarrow10\rightarrow10\rightarrow10\rightarrow500$
& 0.386 (0.008) & 0.770 (0.027)
& 0.461 (0.005) & 0.801 (0.017) & \textbf{0.837 (0.031)} \\

\bottomrule
\end{tabular}
}

\endgroup

\vspace{2pt}
\begin{minipage}{0.96\linewidth}
\footnotesize
\textit{Note:}
All numerical results are mean adjusted Rand indices, with standard
deviations reported in parentheses. Results are summarized over five fitting
seeds. The best result for each setting is shown in bold.
\end{minipage}
\end{table}

\subsubsection{Posterior co-clustering and pairwise uncertainty}\label{sec:sim-unc}

Posterior co-clustering matrices for Simulation~I with $p=200$, $\gamma=0.4$ and $K\in\{4,6,8\}$ are provided in Supplementary Section~S2.2. For $K=4$ they show four distinct blocks with high within-block and low between-block probabilities. For $K=6$, and more so $K=8$, blocks are less distinct and intermediate probabilities appear both within and between groups, consistent with the lower ARI in these settings.

\subsubsection{Selecting the number of clusters}\label{sec:sim-K}
For each true $K\in\{4,6,8\}$ in the setting with $n=800$, $p=200$ and $\gamma=0.4$, we fitted
GB Cluster--Joint at a range of candidate values and computed $Q_K$ in Equation \eqref{eq:QK} as
described in Section~\ref{sec:choose-K} (Supplementary Section~S3,
Figure~S2). For true $K=4$ and $K=6$, $Q_K$ dropped sharply up
to the true value and flattened thereafter, giving a clear elbow at the truth. For true $K=8$,
the curve began to flatten at the candidate value 6, with no clear change at 8. 
This is consistent with the simulation design, where the separation between adjacent centres decreases from $\delta_{6,0.4}\approx 2.8$ to $\delta_{8,0.4}\approx 2.2$, leading to greater overlap between neighbouring clusters when $K=8$.

\section{Real-Data Applications}
\subsection{Datasets and protocol}
We compare the performance of the five methods by using four public benchmarks spanning image, wearable-sensor, EEG and speech features (Table~\ref{tab:benchmark_summary}). For MNIST \citep{lecun1998gradient} we use a class-balanced subset of 2,000 images (200 per digit), representing the moderate-$n$, large-$p$ regime.

UCI HAR \citep{reyesortiz2013har} contains 561 features derived from smartphone accelerometer and gyroscope signals (training set). The Epileptic Seizure Recognition data \citep{wu2017epileptic,andrzejak2001indications} contain 178 EEG features per one-second segment; the five original classes are collapsed into seizure versus non-seizure, so the seizure class forms 20\% of the observations. ISOLET \citep{cole1991isolet} contains 617 acoustic features for 26 spoken letters. 

\begin{table}[htbp]
\centering
\caption{Summary of benchmark datasets used in the real-data applications.}
\label{tab:benchmark_summary}
\small
\begin{tabular}{lccc}
\toprule
Dataset & Sample size & Input dimension & Classes \\
\midrule
MNIST (Balanced Subset)      & 2,000 & 784 & 10 \\
UCI HAR (Training set)   &  7,352 & 561 & 6 \\
Epileptic     & 11,500 & 178 & 2 \\
ISOLET     & 7,797 & 617 & 26 \\
\bottomrule
\end{tabular}
\end{table}

\paragraph{Implementation details}
Before model fitting, MNIST pixel intensities were scaled to \([0,1]\), whereas
the features of UCI HAR, Epileptic, and ISOLET were
standardized coordinate-wise. For the performance reported, \(K\) was fixed at the number of
reference classes. Complete details of the network architectures, preprocessing
procedures and optimization settings are provided in Supplementary
Section~S1.

\subsection{Clustering Performance}

Clustering performance was evaluated using the adjusted Rand index
(ARI) and normalized mutual information (NMI)
\citep{vinh2010information}.
Table~\ref{tab:benchmark_performance} summarizes the results.
\textit{GB Cluster--Joint} achieved the highest mean ARI and NMI
on MNIST, UCI HAR, and ISOLET.
A notable pattern observed is that joint training delivers further gains
when autoencoder preprocessing alone produces modest improvements.
For ISOLET, preprocessing increased mean ARI from 0.467 to 0.503,
while joint training increased it further to 0.579.
The additional gain from joint refinement (0.076) exceeded
the gain from preprocessing (0.036). For Epileptic, most of the improvement came from autoencoder
preprocessing, with GB Cluster--TwoStage performing slightly
better than \textit{GB Cluster--Joint}. Visualizations of the observed and learned representations
for the benchmark datasets are provided in Supplementary
Section~S4.
\begin{table}[htbp]
\centering
\caption{Clustering performance of the proposed
\textit{GB Cluster--Joint} framework and baseline methods on the benchmark
datasets.}
\label{tab:benchmark_performance}

\begin{tabular}{llcc}
\toprule
Dataset & Method & ARI & NMI \\
\midrule

\multirow{5}{*}{MNIST}
& GB Cluster--Raw
& 0.329 (0.031) & 0.421 (0.039) \\
& \(K\)-means
& 0.322 (0.012) & 0.450 (0.013) \\
& GB Cluster--TwoStage
& 0.414 (0.038) & 0.598 (0.030) \\
& AE + \(K\)-means
& 0.421 (0.041) & 0.605 (0.045) \\
& \textit{GB Cluster--Joint}
& \textbf{0.484 (0.081)} & \textbf{0.655 (0.055)} \\
\midrule

\multirow{5}{*}{UCI HAR}
& GB Cluster--Raw
& 0.461 (0.012) & 0.618 (0.018) \\
& \(K\)-means
& 0.456 (0.001) & 0.584 (0.001) \\
& GB Cluster--TwoStage
& 0.500 (0.011) & 0.652 (0.013) \\
& AE + \(K\)-means
& 0.481 (0.019) & 0.630 (0.016) \\
& \textit{GB Cluster--Joint}
& \textbf{0.538 (0.008)} & \textbf{0.731 (0.011)} \\
\midrule

\multirow{5}{*}{Epileptic}
& GB Cluster--Raw
& 0.013 (0.009) & 0.015 (0.005) \\
& \(K\)-means
& 0.014 (0.002) & 0.003 (0.001) \\
& GB Cluster--TwoStage
& \textbf{0.729 (0.008)} & \textbf{0.530 (0.007)} \\
& AE + \(K\)-means
& 0.720 (0.010) & 0.521 (0.009) \\
& \textit{GB Cluster--Joint}
& 0.720 (0.006) & 0.517 (0.007) \\
\midrule

\multirow{5}{*}{ISOLET}
& GB Cluster--Raw
& 0.467 (0.007) & 0.713 (0.010) \\
& \(K\)-means
& 0.469 (0.021) & 0.710 (0.011) \\
& GB Cluster--TwoStage
& 0.503 (0.013) & 0.728 (0.009) \\
& AE + \(K\)-means
& 0.500 (0.026) & 0.721 (0.013) \\
& \textit{GB Cluster--Joint}
& \textbf{0.579 (0.026)} & \textbf{0.753 (0.016)} \\
\bottomrule
\end{tabular}

\vspace{2pt}
\begin{minipage}{0.82\linewidth}
\footnotesize
\textit{Note:}
Entries report mean performance across five runs with different
random seeds, with sample standard deviations in parentheses.
The best result for each dataset and metric is shown in bold.
\end{minipage}
\end{table}
\subsection{Posterior Uncertainty}

\begin{figure}[!htbp]
\centering
\includegraphics[width=0.95\linewidth]
{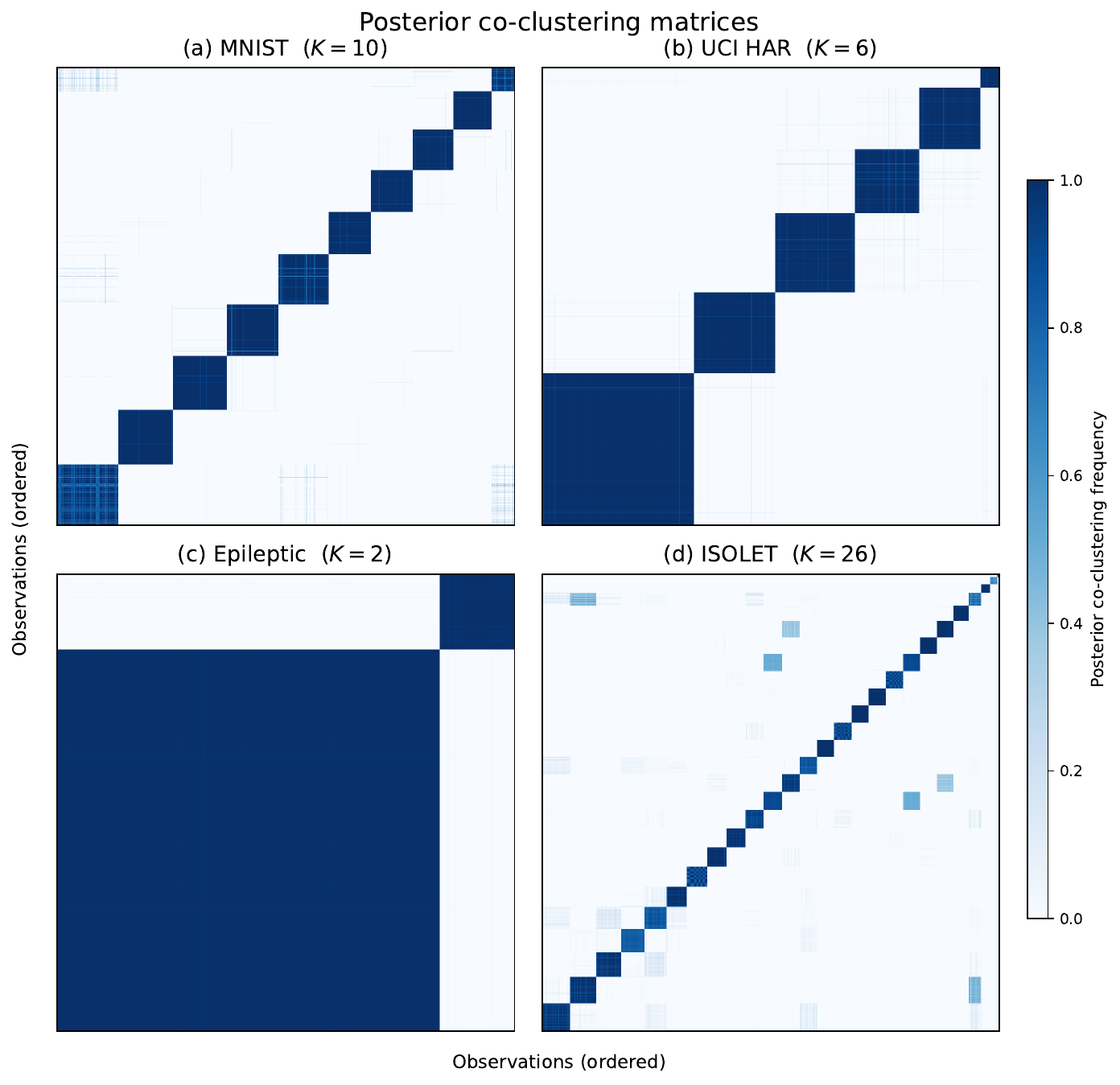}
\caption{
Posterior co-clustering matrices for (a) MNIST, (b) UCI HAR,
(c) Epileptic, and (d) ISOLET, shown on a common scale from
zero to one. Darker entries indicate that a pair of observations was assigned to the same cluster more frequently.
}
\label{fig:supp-benchmark-coclustering}
\end{figure}

The posterior co-clustering matrices in
Figure~\ref{fig:supp-benchmark-coclustering}
summarize allocation uncertainty conditional on the fitted
representations and selected numbers of clusters. For UCI HAR and Epileptic, co-clustering probabilities are
predominantly close to zero or one, indicating little variation
in pairwise co-membership across the retained draws. MNIST also shows a recognizable block structure, although
some observation pairs have intermediate co-clustering
probabilities. ISOLET displays the most heterogeneous pattern,
with several less sharply defined diagonal blocks and
localized off-diagonal associations. Overall, the generally low
uncertainty in these applications is consistent with the
real-data findings of \citet{rigon2023generalized}.

\begin{figure}[!htbp]
\centering
\includegraphics[width=0.95\linewidth]
{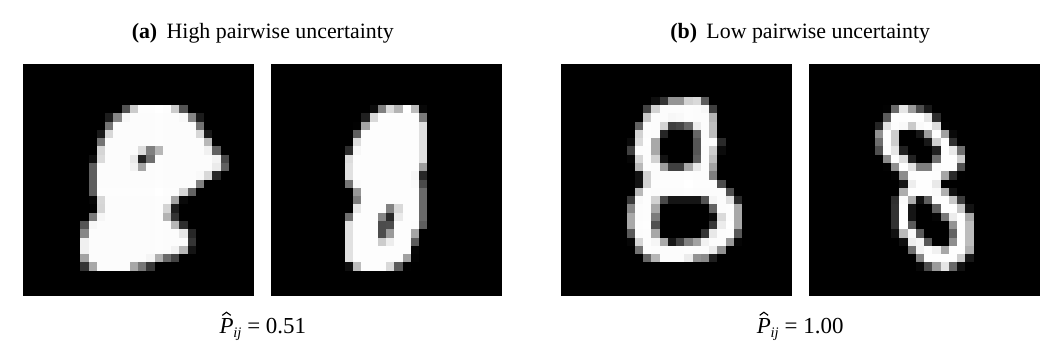}
\caption{
Illustrative MNIST image pairs with (a) Relatively high
(\(\widehat P_{ij}=0.51\)) and (b) Relatively low
(\(\widehat P_{ij}=1.00\)) pairwise posterior uncertainty.
All four images have reference label~8. 
}
\label{fig:mnist_sample}
\end{figure}

\noindent
Figure~\ref{fig:mnist_sample} illustrates this variation using
two pairs of MNIST images (all with reference label~8).
For the pair in panel~(a), the estimated co-clustering
probability of \(\widehat P_{ij}=0.51\) indicates
co-assignment in approximately half of the retained draws. In panel~(b), both digits have
more clearly defined upper and lower loops, with an estimated
co-clustering probability of $\widehat P_{ij}=1.00$ at the
displayed precision. Within the same fitted model, the pair
in panel~(a) therefore exhibits greater ambiguity about
co-membership than the pair in panel~(b).

\subsection{Selecting the number of clusters and sensitivity to \texorpdfstring{$r_0$}{r0}}

Following the same elbow heuristic, we refitted GB Cluster--Joint over a range of candidate $K$ and plotted the normalized clustering loss $Q_K$
(Supplementary Section~S3, Figure~S3).
For Epileptic, a clear elbow was observed at $K=2$, matching
the two reference classes. For HAR, the most pronounced change
occurred around $K=5$, compared with six reference classes,
whereas for MNIST the curve began to flatten around $K=8$,
compared with ten reference classes. For ISOLET, which contains
26 reference classes, the elbow was less sharply defined,
suggesting a broader range of candidate values of $K$.

We also examined sensitivity to \(r_0\). The results were relatively
stable over the values considered for all four benchmark datasets.
Detailed results are provided in the Supplementary Material (Table~S4).

\section{Discussion}

In this paper, we have extended the generalized Bayes clustering
framework to incorporate representation learning, allowing reconstruction
and clustering losses to jointly guide the learned representation
while retaining probabilistic inference over partitions.
Compared with generalized Bayes clustering applied in the original
feature space \citep{rigon2023generalized}, the proposed method achieved
higher agreement with the generating or reference partitions across
the simulation studies and benchmark applications. The improvements were
more pronounced when clustering in the observed feature space provided
limited partition recovery. These results provide empirical support for
incorporating representation learning into generalized Bayes clustering
when a fixed discrepancy alone in the original high-dimensional space does not
adequately capture the cluster structure of interest.
The framework also suggests several natural extensions. First, the posterior
co-clustering probabilities are computed conditional on the fitted
representation, which is held fixed during the final sampling stage.
Propagating uncertainty from representation learning, for example by sampling
the network parameters or averaging over multiple fitted representations, would
yield summaries that more fully reflect all sources of uncertainty. Second,
the number of clusters is fixed in each fit, so the co-clustering probabilities
describe allocation uncertainty at a given resolution. Placing a prior on $K$ would allow uncertainty in the number of clusters to be quantified jointly with
the allocation. Finally, our empirical study focused on relatively simple autoencoder architectures. Because the
framework accommodates any differentiable
encoder, it extends directly to more specialized representation models. Exploring these settings is a
promising direction for future work.

\bibliographystyle{elsarticle-harv}
\bibliography{references}
\end{document}

\maketitle
\vspace{-1.5em}

This supplement provides implementation details and additional results supporting the main
paper. Section~\ref{sec:supp-implementation} gives the autoencoder
architectures and training settings for all analyses. Section~\ref{sec:supp-simulation1}
reports additional Simulation~I results, including posterior co-clustering matrices.
Section~\ref{sec:supp-k-curves} presents the elbow plots used to select the number of clusters,
Section~\ref{sec:supp-benchmark-representations} visualizes the observed and learned
representations of the benchmark datasets, and Section~\ref{sec:supp-r0-sensitivity} examines
the sensitivity of \GBJoint\ to \(r_0\).

% ======================================================================
\section{Implementation Details}
\label{sec:supp-implementation}

\subsection{Simulation studies}
\label{sec:supp-simulation-implementation}
In Simulation~I settings, the autoencoder architecture was \(p\to d\to p\), with latent dimension
\(d=4\) for \(p\in\{50,200\}\), \(d=6\) for \(p=500\) and \(d=8\) for \(p=1000\). The
autoencoder was pretrained for 100 epochs with learning rate \(5\times10^{-3}\), and \GBJoint\
was then run for 100 joint iterations with learning rate \(10^{-3}\); the batch size was 64.

In all Simulation~II settings, the encoder maps the \(p\) observed coordinates to a hidden layer of width
32, followed by layer normalization and a GELU activation, and then to a latent representation
of dimension \(d=4\). The decoder maps the latent representation to a hidden layer of width 32
with a GELU activation and then to a linear output of dimension \(p\). No dropout was used.
The autoencoder was pretrained for 200 epochs, and \GBJoint\ was then run for 100 joint
iterations, with a learning rate of \(10^{-3}\) in both stages and a batch size of 128.

\subsection{Benchmark datasets}
\label{sec:real-implementation}

\paragraph{Data preprocessing.}
For MNIST, the input pixel intensities were divided by 255
to scale them to \([0,1]\). For UCI HAR and Epileptic, each feature was centred and
scaled using its mean and standard deviation over the
7,352 and 11,500 analysed observations, respectively. For ISOLET, feature means and sample standard deviations
were computed from the original training set of 6,238
observations and the resulting transformation was applied
to all 7,797 observations.

\paragraph{Autoencoder architectures and training settings.}
The autoencoder architectures and training settings are summarized
in Table~\ref{tab:supp-real-ae-settings}. MNIST used a convolutional
autoencoder, whereas UCI HAR, Epileptic, and ISOLET used fully
connected autoencoders with symmetric encoder and decoder widths.
ReLU activations were used in the hidden layers. The reconstruction output used a sigmoid
activation for MNIST and a linear activation for the other datasets. All autoencoders were pretrained for 100 epochs using Adam,
with the learning rates and batch sizes reported in
Table~\ref{tab:supp-real-ae-settings}. AE + \(K\)-means and \GBTwo\ both operated
directly on the pretrained representations, using \(K\)-means and generalized Bayes clustering,
respectively. \GBJoint\ used the same autoencoder
configuration and continued updating the pretrained network
under the joint loss.

\begin{table}[htbp]
\centering
\caption{Autoencoder architectures and training settings for the
real-data applications.}
\label{tab:supp-real-ae-settings}
\small
\setlength{\tabcolsep}{4pt}
\begin{tabular}{llccccc}
\toprule
Dataset & Encoder & \(d\) & Batch size
& \shortstack{Pretraining\\learning rate}
& \shortstack{Joint\\learning rate}
& \shortstack{Joint\\iterations} \\
\midrule
MNIST
& Conv16--Conv32--FC10
& 10 & 128 & \(5\times10^{-4}\) & \(10^{-4}\) & 100 \\
UCI HAR
& 561--256--64--5
& 5 & 128 & \(10^{-4}\) & \(10^{-4}\) & 100 \\
Epileptic
& 178--128--32--10
& 10 & 128 & \(10^{-4}\) & \(10^{-4}\) & 100 \\
ISOLET
& 617--256--64--16
& 16 & 128 & \(10^{-4}\) & \(10^{-4}\) & 100 \\
\bottomrule
\end{tabular}

\smallskip
\begin{minipage}{\linewidth}
\footnotesize
\textit{Note:}
\(d\) denotes the latent dimension. Fully connected decoders
mirror the encoder widths. For MNIST, all convolutions and transposed convolutions use
a \(4\times4\) kernel, stride 2, and padding 1, with output
padding 0 for transposed convolutions.
The decoder maps the latent vector to 1,568 units, reshapes
them into \(32\times7\times7\) feature maps, and applies
two transposed convolutions with 16 and 1 output channels.
ReLU follows the decoder's fully connected layer and first
transposed convolution; the output uses a sigmoid activation.
\end{minipage}
\end{table}

% ======================================================================
\section{Additional Results for Simulation I}
\label{sec:supp-simulation1}

\subsection{Clustering performance for \texorpdfstring{\(K=6\) and \(K=8\)}{K=6 and K=8}}
\label{sec:supp-sim1-performance}
Tables~\ref{tab:supp-sim1-k6} and~\ref{tab:supp-sim1-k8} report Simulation~I results for $K=6$ and
$K=8$ under the same design and protocol as Table~1 of the main paper. At a given $\gamma$,
adjacent cluster centres are closer than for $K=4$, which makes clustering harder. GB Cluster--Joint had the highest or joint-highest mean ARI in all but two configurations, both
at $p=50$ and within 0.021 of the best method. Its advantage over GB Cluster--TwoStage was
smaller than for $K=4$ and did not grow steadily with $p$; for $K=8$ and $p=1000$, all methods
had mean ARI below 0.40.
\begin{table}[!ht]
\centering
\caption{
Mean adjusted Rand index (ARI) across five runs with different random seeds for the reported \((p,\gamma)\) configurations in Simulation~I
with \(n=800\) and \(K=6\). Sample standard deviations are shown in
parentheses.
}
\label{tab:supp-sim1-k6}

\small
\begin{tabular}{lrrrrr}
\toprule
Method & \(p\) & \(\gamma=0.3\) & \(\gamma=0.4\)
& \(\gamma=0.5\) & \(\gamma=0.6\) \\
\midrule

\GBRaw   & 50   & 0.362 (0.006) & 0.553 (0.003) & 0.791 (0.001) & 0.918 (0.006) \\
\GBRaw   & 200  & 0.244 (0.009) & 0.440 (0.053) & 0.674 (0.001) & 0.847 (0.008) \\
\GBRaw   & 500  & 0.153 (0.007) & 0.358 (0.026) & 0.506 (0.055) & 0.717 (0.002) \\
\GBRaw   & 1000 & 0.092 (0.009) & 0.253 (0.004) & 0.361 (0.021) & 0.553 (0.015) \\

\addlinespace

\(K\)-means & 50   & 0.323 (0.022) & 0.571 (0.033) & 0.786 (0.020) & 0.917 (0.017) \\
\(K\)-means & 200  & 0.248 (0.020) & 0.417 (0.018) & 0.628 (0.046) & 0.846 (0.052) \\
\(K\)-means & 500  & 0.143 (0.011) & 0.324 (0.021) & 0.496 (0.026) & 0.693 (0.041) \\
\(K\)-means & 1000 & 0.089 (0.004) & 0.254 (0.010) & 0.331 (0.058) & 0.531 (0.041) \\

\addlinespace

\GBTwo   & 50   & \textbf{0.411 (0.017)} & 0.611 (0.009) & 0.801 (0.008) & 0.910 (0.005) \\
\GBTwo   & 200  & 0.300 (0.043) & 0.558 (0.090) & 0.803 (0.031) & 0.898 (0.005) \\
\GBTwo   & 500  & 0.178 (0.012) & 0.410 (0.026) & 0.686 (0.048) & 0.857 (0.017) \\
\GBTwo   & 1000 & 0.077 (0.006) & 0.286 (0.009) & 0.393 (0.016) & 0.627 (0.058) \\

\addlinespace

AE + \(K\)-means & 50   & 0.380 (0.052) & \textbf{0.633 (0.034)} & 0.807 (0.014) & 0.911 (0.010) \\
AE + \(K\)-means & 200  & 0.282 (0.050) & 0.540 (0.063) & 0.771 (0.017) & 0.894 (0.014) \\
AE + \(K\)-means & 500  & 0.169 (0.017) & 0.380 (0.066) & 0.662 (0.017) & 0.805 (0.059) \\
AE + \(K\)-means & 1000 & 0.068 (0.004) & 0.265 (0.018) & 0.374 (0.022) & 0.629 (0.039) \\

\addlinespace

\GBJoint & 50   & 0.409 (0.008) & 0.612 (0.011) & \textbf{0.835 (0.008)} & \textbf{0.931 (0.016)} \\
\GBJoint & 200  & \textbf{0.324 (0.010)} & \textbf{0.608 (0.011)} & \textbf{0.829 (0.012)} & \textbf{0.925 (0.009)} \\
\GBJoint & 500  & \textbf{0.200 (0.004)} & \textbf{0.445 (0.009)} & \textbf{0.713 (0.010)} & \textbf{0.888 (0.011)} \\
\GBJoint & 1000 & \textbf{0.109 (0.002)} & \textbf{0.332 (0.003)} & \textbf{0.446 (0.015)} & \textbf{0.654 (0.018)} \\

\bottomrule
\end{tabular}
\par\smallskip
\begin{minipage}{\linewidth}
\raggedright
\footnotesize
\textit{Note: The best result for each setting is shown in bold.}
\end{minipage}
\end{table}

\begin{table}[!ht]
\centering
\caption{
Mean adjusted Rand index (ARI) across five runs with different random seeds for the reported \((p,\gamma)\) configurations in Simulation~I
with \(n=800\) and \(K=8\). Sample standard deviations are shown in
parentheses.
}
\label{tab:supp-sim1-k8}

\small
\begin{tabular}{lrrrrr}
\toprule
Method & \(p\) & \(\gamma=0.3\) & \(\gamma=0.4\)
& \(\gamma=0.5\) & \(\gamma=0.6\) \\
\midrule

\GBRaw   & 50   & 0.241 (0.030) & 0.402 (0.021) & 0.554 (0.021) & 0.735 (0.029) \\
\GBRaw   & 200  & 0.155 (0.024) & 0.266 (0.030) & 0.351 (0.023) & 0.541 (0.009) \\
\GBRaw   & 500  & 0.116 (0.009) & 0.222 (0.010) & 0.320 (0.011) & 0.390 (0.008) \\
\GBRaw   & 1000 & 0.055 (0.021) & 0.173 (0.028) & 0.263 (0.019) & 0.335 (0.007) \\

\addlinespace

\(K\)-means & 50   & 0.221 (0.008) & 0.350 (0.011) & 0.499 (0.049) & 0.700 (0.055) \\
\(K\)-means & 200  & 0.121 (0.016) & 0.245 (0.010) & 0.361 (0.028) & 0.531 (0.018) \\
\(K\)-means & 500  & 0.101 (0.009) & 0.195 (0.032) & 0.316 (0.013) & 0.372 (0.027) \\
\(K\)-means & 1000 & 0.047 (0.002) & 0.128 (0.044) & 0.235 (0.035) & 0.316 (0.022) \\

\addlinespace

\GBTwo   & 50   & \textbf{0.268 (0.014)} & 0.459 (0.020) & 0.604 (0.008) & 0.754 (0.012) \\
\GBTwo   & 200  & 0.207 (0.009) & 0.356 (0.011) & 0.420 (0.003) & 0.707 (0.014) \\
\GBTwo   & 500  & 0.082 (0.009) & 0.277 (0.005) & 0.354 (0.008) & 0.614 (0.010) \\
\GBTwo   & 1000 & 0.063 (0.002) & 0.194 (0.005) & 0.279 (0.008) & 0.381 (0.005) \\

\addlinespace

AE + \(K\)-means & 50   & 0.250 (0.034) & 0.393 (0.026) & 0.570 (0.083) & 0.747 (0.025) \\
AE + \(K\)-means & 200  & 0.167 (0.049) & 0.327 (0.028) & 0.419 (0.057) & 0.688 (0.043) \\
AE + \(K\)-means & 500  & 0.095 (0.011) & 0.229 (0.022) & 0.292 (0.070) & 0.541 (0.064) \\
AE + \(K\)-means & 1000 & 0.055 (0.003) & 0.172 (0.007) & 0.259 (0.016) & 0.356 (0.057) \\

\addlinespace

\GBJoint & 50   & \textbf{0.268 (0.012)} & \textbf{0.470 (0.009)} & \textbf{0.625 (0.018)} & \textbf{0.770 (0.018)} \\
\GBJoint & 200  & \textbf{0.239 (0.030)} & \textbf{0.368 (0.019)} & \textbf{0.502 (0.018)} & \textbf{0.757 (0.004)} \\
\GBJoint & 500  & \textbf{0.130 (0.009)} & \textbf{0.297 (0.007)} & \textbf{0.403 (0.010)} & \textbf{0.635 (0.007)} \\
\GBJoint & 1000 & \textbf{0.089 (0.004)} & \textbf{0.201 (0.010)} & \textbf{0.297 (0.024)} & \textbf{0.388 (0.008)} \\

\bottomrule
\end{tabular}
\par\smallskip
\begin{minipage}{\linewidth}
\raggedright
\footnotesize
\textit{Note: The best result for each setting is shown in bold.}
\end{minipage}
\end{table}

\FloatBarrier

\subsection{Posterior co-clustering uncertainty}
\label{sec:supp-uncertainty}
% Retain former heading labels as aliases for existing main-text references.
\label{sec:supp-oracle-assessment}
\label{sec:supp-psm-definition}
\label{sec:supp-simulation-uncertainty}
\label{sec:supp-coclustering}
% The former equation (S1) and its alias eq:supp-coclustering were removed: they repeated
% the definition of the co-clustering probabilities in Section 3.4 of the main paper.

\begin{figure}[h]
\centering
\includegraphics[width=\linewidth]
{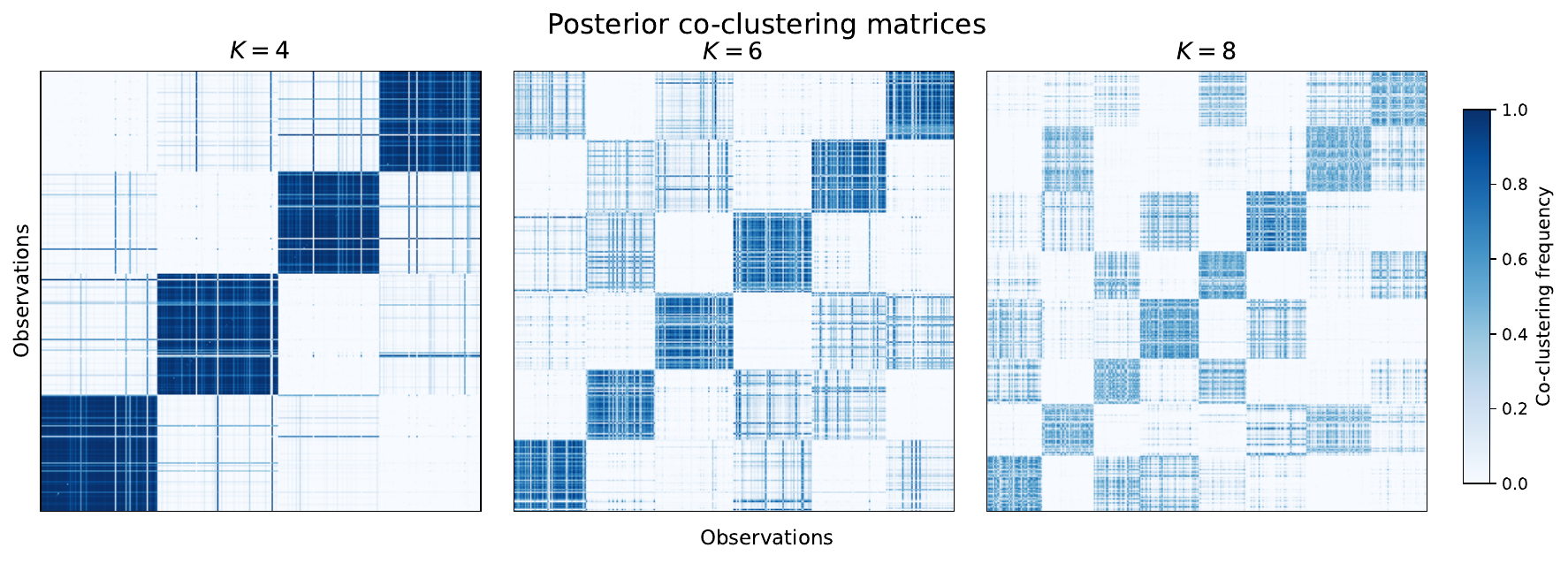}
\caption{
Posterior co-clustering matrices obtained using \GBJoint\
for Simulation~I with \(p=200\), \(\gamma=0.4\).
The panels correspond to \(K=4,6,8\) from left to right. Darker shades indicate higher estimated posterior
co-clustering probabilities.
}
\label{fig:supp-sim1-coclustering}
\end{figure}

Figure~\ref{fig:supp-sim1-coclustering} displays the posterior co-clustering matrices, whose
entries \(\widehat P_{ij}\) are defined in Section~3.4 of the main paper, for \GBJoint\ in
Simulation~I with \(n=800\), \(p=200\), \(\gamma=0.4\) and \(K=4,6,8\).

% ======================================================================
\section{Elbow Plots for Selecting the Number of Clusters}
\label{sec:supp-k-curves}
\begin{figure}[!htbp]
\centering
\includegraphics[width=\linewidth]
{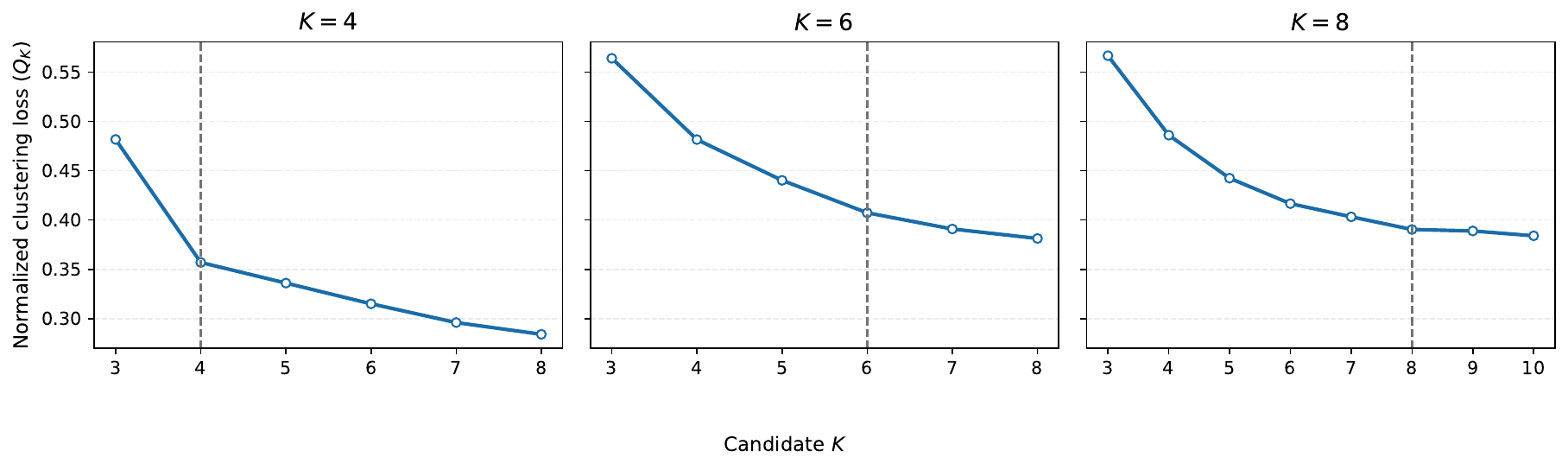}
\caption{
Normalized clustering loss \(Q_K\) against candidate \(K\) for
\GBJoint\ in Simulation~I with \(n=800\), \(p=200\) and \(\gamma=0.4\), for true
\(K=4,6,8\) (left to right). Dashed vertical lines mark the true number of clusters.
}
\label{fig:supp-sim1-k-curves}
\end{figure}

\begin{figure}[!htbp]
\centering
\includegraphics[width=0.96\linewidth]
{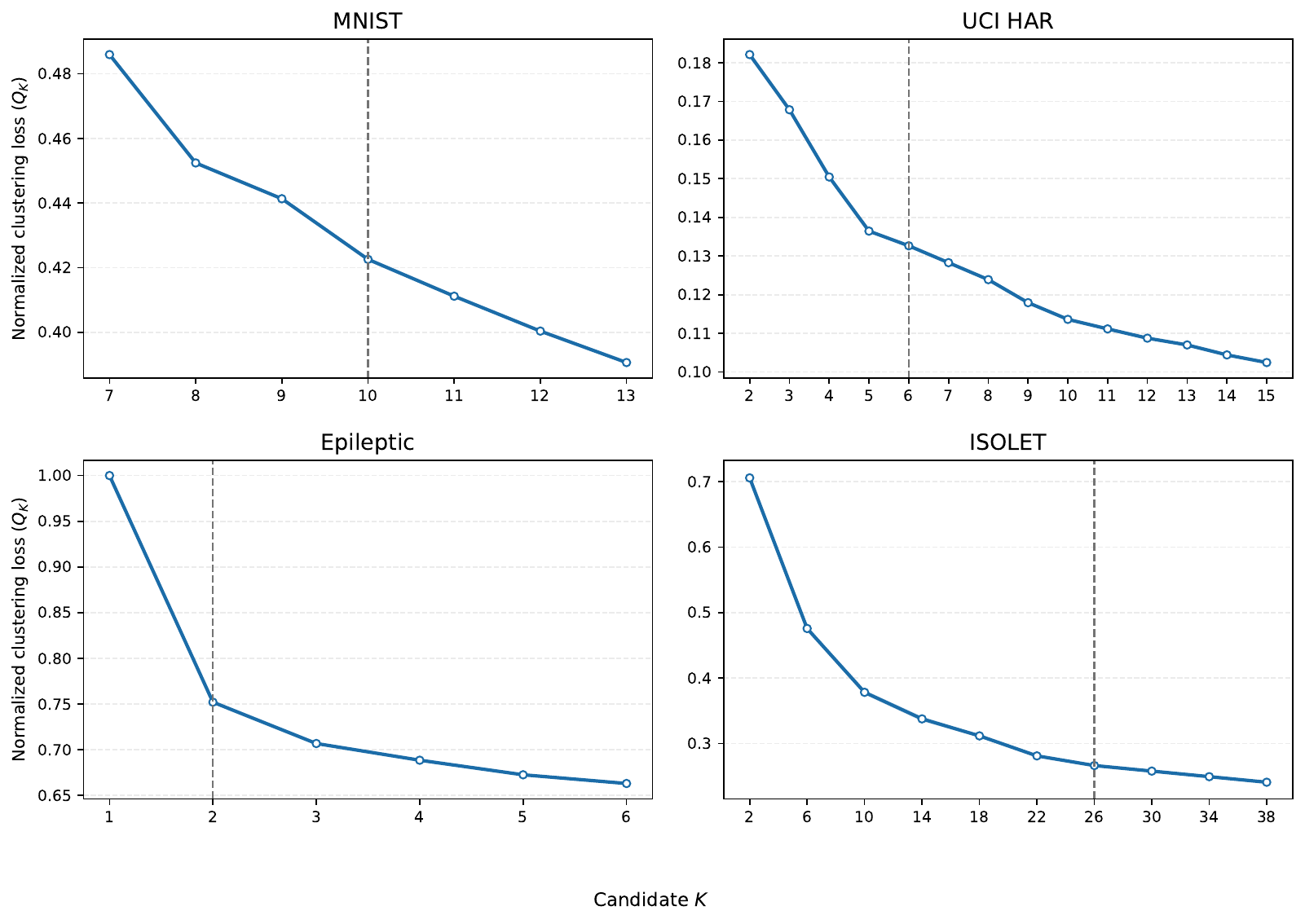}
\caption{
Normalized clustering loss \(Q_K\) against candidate \(K\) for
\GBJoint\ on the benchmark datasets. Dashed vertical lines mark the number of reference
classes.
}
\label{fig:supp-benchmark-k-curves}
\end{figure}

% Q_K is defined in Section 3.5 of the main paper; update "(22)" if its number differs.
Figures~\ref{fig:supp-sim1-k-curves} and~\ref{fig:supp-benchmark-k-curves} plot the normalized
clustering loss \(Q_K\) against candidate values of \(K\) for
\GBJoint, in selected Simulation~I settings and on the benchmark datasets, respectively.

% ======================================================================
\section{Additional Benchmark Representation Visualizations}
\label{sec:supp-benchmark-representations}
% TODO (not shown in PDF): report t-SNE settings (perplexity, seed) and whether all
% observations are plotted. Figures S5 and S6 report explained variance differently
% (combined in panel titles vs per axis); consider harmonizing.

\begin{figure}[!htbp]
\centering
\includegraphics[width=\linewidth]
{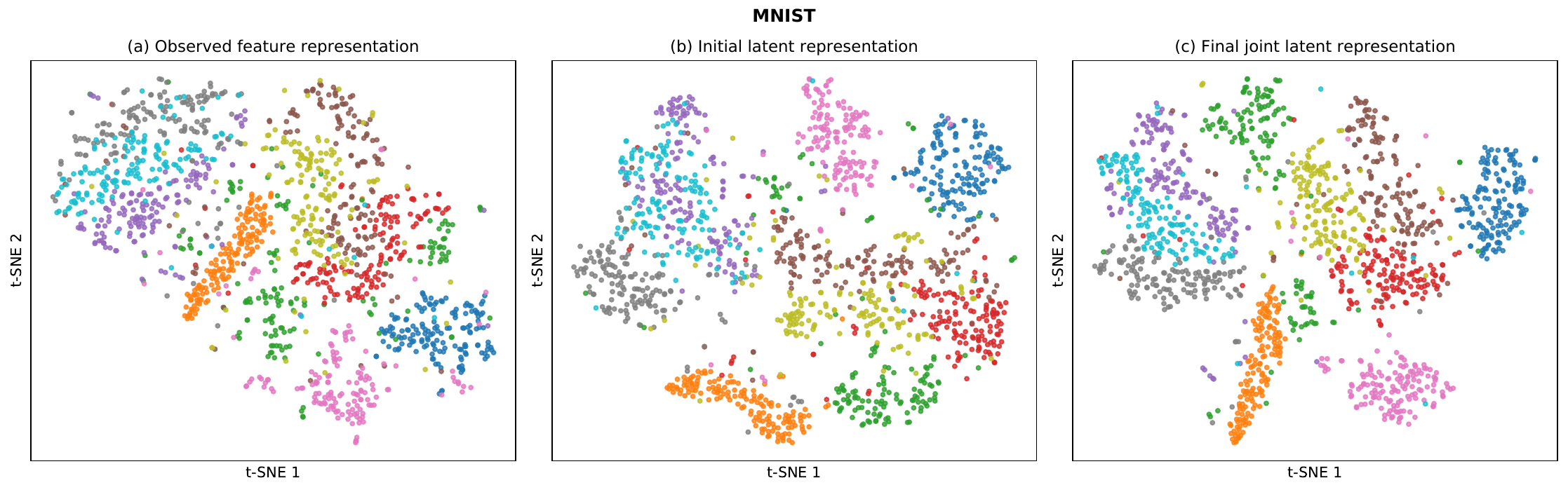}
\caption{
MNIST: t-SNE visualizations of the observed feature representation
(left), initial latent representation (middle), and final joint
latent representation (right). 
}
\label{fig:supp-mnist-representations}
\end{figure}

\begin{figure}[!htbp]
\centering
\includegraphics[width=\linewidth]
{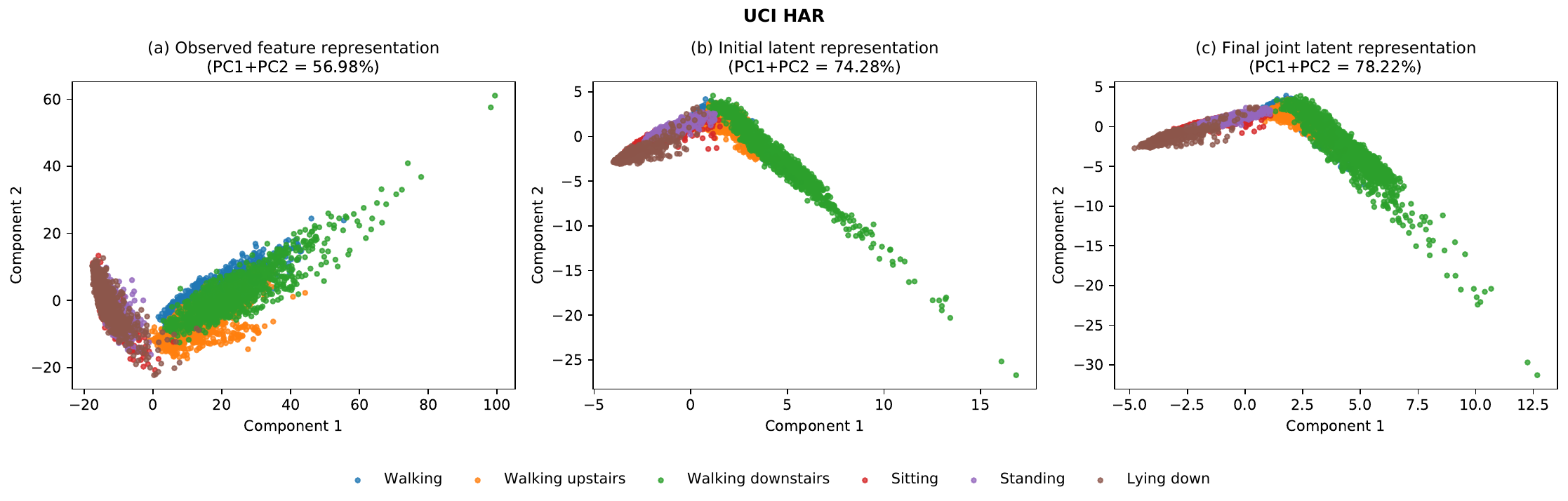}
\caption{
UCI HAR: PCA visualizations of the observed feature representation
(left), initial latent representation (middle), and final joint
latent representation (right). Percentages below the panel titles
give the combined variance explained by the two displayed
principal components.
}
\label{fig:supp-har-representations}
\end{figure}

\begin{figure}[!htbp]
\centering
\includegraphics[width=\linewidth]
{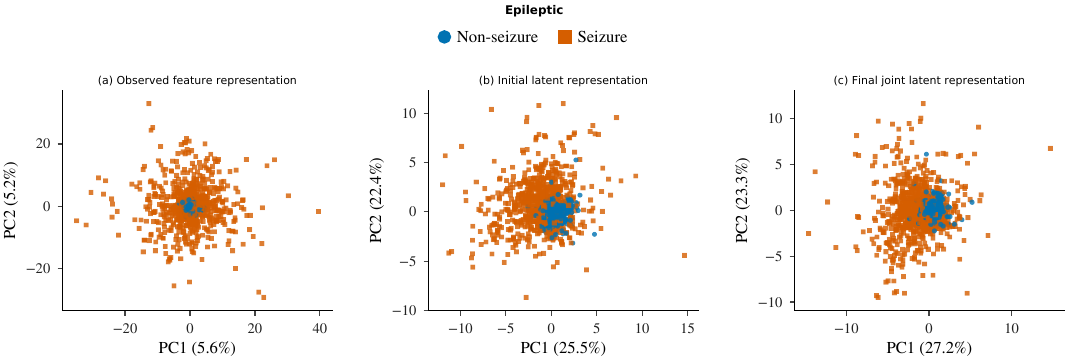}
\caption{
Epileptic: PCA visualizations of the observed feature representation
(left), initial latent representation (middle), and final joint
latent representation (right). Percentages on the axes give the
variance explained by each displayed principal component.
}
\label{fig:supp-epileptic-representations}
\end{figure}

\begin{figure}[!htbp]
\centering
\includegraphics[width=\linewidth]
{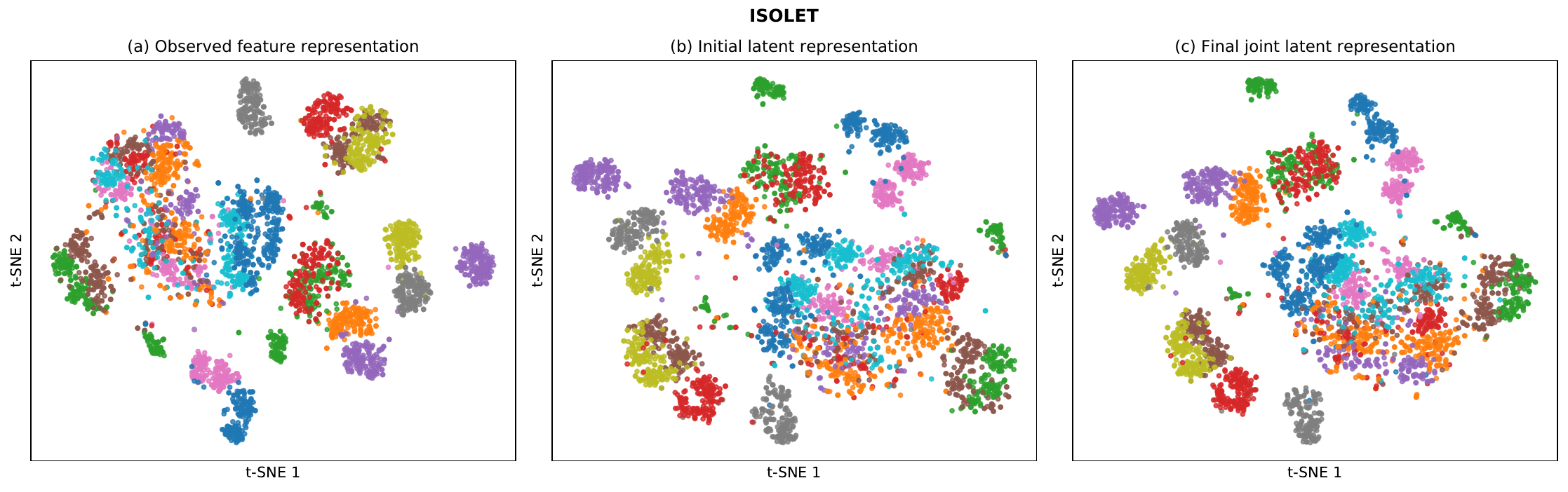}
\caption{
ISOLET: t-SNE visualizations of the observed feature representation
(left), initial latent representation (middle), and final joint
latent representation (right). 
}
\label{fig:supp-isolet-representations}
\end{figure}

Figures~\ref{fig:supp-mnist-representations}--\ref{fig:supp-isolet-representations}
compare the observed feature representations (left), initial latent
representations obtained by reconstruction-only pretraining (middle),
and final joint latent representations (right) for MNIST, UCI HAR,
Epileptic, and ISOLET, respectively. Colours denote different reference classes. PCA is used for
UCI HAR and Epileptic, whereas t-SNE is used for MNIST and ISOLET.

% ======================================================================
\section{Sensitivity to \texorpdfstring{\(r_0\)}{r0}}
\label{sec:supp-r0-sensitivity}

Table~\ref{tab:sensitivity-r0} examines sensitivity to the target ratio \(r_0\) used to
calibrate \(\lambda\); larger \(r_0\) gives more weight to
reconstruction. We fit \GBJoint\ with \(r_0\in\{0.5,1,2\}\), all other settings unchanged, in
Simulation~I (\(n=800\), \(K=4\), \(\gamma=0.4\)) and on the four benchmark datasets. Except on Epileptic, \(r_0=0.5\) and
\(r_0=1\) gave similar results, differing by at most 0.033. Setting \(r_0=2\) lowered the ARI
by 0.05 to 0.07 in Simulation~I and by 0.07 on ISOLET. Epileptic was the most sensitive dataset: relative to \(r_0=1\), the mean ARI was 0.092 lower
for \(r_0=0.5\) and 0.047 lower for \(r_0=2\), and the standard deviation across seeds increased
from 0.006 to 0.078 and 0.059, respectively.

\begin{table}[htbp]
\centering
\caption{Sensitivity of \GBJoint\ to the target gradient ratio \(r_0\). Entries report the mean adjusted Rand index (ARI) with respect to the generating
(Simulation~I) or reference (benchmark) labels across five random seeds, with sample standard
deviations in parentheses.}
\label{tab:sensitivity-r0}
\begin{tabular}{lccc}
\toprule
Setting & \(r_0=0.5\) & \(r_0=1\) & \(r_0=2\) \\
\midrule
\multicolumn{4}{l}{\textit{Simulation I}
(\(n=800,\ K=4,\ \gamma=0.4\))} \\
\(p=50\)   & 0.887 (0.015) & 0.900 (0.041) & 0.849 (0.009) \\
\(p=200\)  & 0.816 (0.044) & 0.849 (0.018) & 0.780 (0.059) \\
\(p=500\)  & 0.808 (0.029) & 0.777 (0.017) & 0.712 (0.061) \\
\(p=1000\) & 0.681 (0.030) & 0.682 (0.038) & 0.622 (0.043) \\
\addlinespace
\multicolumn{4}{l}{\textit{Benchmark datasets}} \\
MNIST     & 0.489 (0.033) & 0.484 (0.081) & 0.484 (0.033) \\
UCI HAR   & 0.559 (0.022) & 0.538 (0.008) & 0.541 (0.013) \\
Epileptic & 0.628 (0.078) & 0.720 (0.006) & 0.673 (0.059) \\
ISOLET    & 0.561 (0.024) & 0.579 (0.026) & 0.505 (0.024) \\
\bottomrule
\end{tabular}
\end{table}